\documentclass{article}
\usepackage{graphicx}
\usepackage{amsmath,amssymb,amsfonts}
\usepackage{parskip}
\usepackage[numbers, square, comma]{natbib}
\usepackage[nottoc,numbib]{tocbibind}
\usepackage{authblk}
\usepackage[margin=1.6in]{geometry}

\begin{document}

\title{Photoinduced twist and untwist of moir\'{e} superlattices}

\author[1,2]{C. J. R. Duncan$^\bigstar$}

\author[3]{A. C. Johnson}

\author[4,5]{I. Maity}

\author[5,6]{A. Rubio}

\author[1]{M. Gordon}

\author[1]{A. C. Bartnik}

\author[1,7]{M Kaemingk}

\author[1,8]{W. H. Li}

\author[1]{M. B. Andorf}

\author[1,9]{C. A. Pennington}

\author[1]{I. V. Bazarov}

\author[10]{M. W. Tate}

\author[11,12]{D. A. Muller}

\author[13]{J. Thom-Levy}

\author[10,14]{S. M. Gruner}

\author[2,3]{A . M. Lindenberg}

\author[1]{J. M. Maxson$^\bigstar$}

\author[15]{F. Liu$^\bigstar$}

\affil[1]{Cornell Laboratory for Accelerator-Based Sciences and Education, Cornell University, Ithaca, NY, USA}

\affil[2]{SLAC National Accelerator Laboratory, Menlo Park, CA, USA}

\affil[3]{Department of Materials Science and Engineering, Stanford University, Stanford, CA, USA}

\affil[4]{Newcastle University, Newcastle, UK}

\affil[5]{Max Planck Institute for the Structure and Dynamics of Matter, Hamburg, Germany}

\affil[6]{Center for Computational Quantum, The Flatiron Institute, New York, NY, USA}

\affil[7]{Los Alamos National Laboratory, Los Alamos, NM, USA}

\affil[8]{Brookhaven National Laboratory, Upton, NY, USA}

\affil[9]{University of California Los Angeles, Los Angeles, CA, USA}

\affil[10]{Laboratory of Atomic and Solid State Physics, Cornell University, Ithaca, NY, USA}

\affil[11]{School of Applied and Engineering Physics, Cornell University, Ithaca, NY, USA}

\affil[12]{Kavli Institute at Cornell for Nanoscale Science, Cornell University, Ithaca, NY, USA}

\affil[13]{Laboratory for Elementary-Particle Physics, Cornell University, Ithaca, NY, USA}

\affil[14]{Cornell High Energy Synchrotron Source (CHESS), Cornell University, Ithaca, NY, USA}

\affil[15]{Department of Chemistry, Stanford University, Stanford, CA, USA}

\date{}
\maketitle
{$^\bigstar$\textit{Corresponding authors}: cjduncan@slac.stanford.edu, jmm586@cornell.edu, \newline fliu10@stanford.edu}
\newpage
\begin{abstract} Two-dimensional moir\'{e} materials are formed by artificially stacking atomically thin monolayers. A wealth of correlated and topological quantum phases can be engineered via precise choice of stacking geometry \cite{cao2018correlated,regan2020mott,xia2025superconductivity}. These designer electronic properties depend crucially on interlayer coupling and atomic registry \cite{yoo2019atomic, xie2022twist}. An important open question is how atomic registry responds on ultrafast timescales to optical excitation and whether the moir\'{e} geometry can be dynamically reconfigured to tune emergent phenomena in real time. Here we show that femtosecond photoexcitation drives a coherent twist-untwist motion of the moir\'{e} superlattice in $2^\circ$ and $57^\circ$ twisted WSe$_2$/MoSe$_2$ heterobilayers, resolved directly by ultrafast electron diffraction. Upon above-band-gap photoexcitation, the moir\'{e} superlattice diffraction features are enhanced within 1 ps and subsequently suppressed several picoseconds after, deviating markedly from typical photoinduced lattice heating. Kinetic diffraction analysis, supported by simulations of the sample dynamics, indicates a peak-to-trough local twist angle modulation of $0.6^\circ$, correlated with a sub-THz frequency moir\'{e} phonon. This motion is driven by ultrafast charge transfer that transiently increases interlayer attraction. Our results could lead to ultrafast control of moir\'{e} periodic lattice distortions and, by extension, the local moiré potential that shapes excitons, polarons, and correlation-driven behaviors.\end{abstract}

\maketitle

\section{Introduction}

Moir\'{e} structures in twisted bilayers have become versatile platforms for engineering a wide range of quantum phenomena, including correlated insulators \cite{cao2018correlated, wang2020correlated, shimazaki2020strongly}, the fractional quantum anomalous Hall effect \cite{Park2023fractional}, generalized Wigner crystals 
\cite{regan2020mott, li2021imaging, zhou2021bilayer},
and superconductivity \cite{xia2025superconductivity, guo2025superconductivity}. The band structure, magnetism, and superconducting properties of these novel materials can be tuned with twist angles and band fillings, strongly influenced by the interplay between the moir\'{e} superlattice structure and electronic potentials \cite{yoo2019atomic, carr2020electronic, xie2022twist, xia2025superconductivity, guo2025superconductivity}.
The spatially varying atomic registry between the constituent layers leads to periodic modulation of interlayer coupling and local stacking energies. Consequently, the moir\'{e} superlattice is not a simple rigid stacking of two layers, but undergoes periodic lattice distortion (PLD): additional local atomic relaxation in each moir\'{e} unit cell, twisted around stacking domain centers \cite{maity2021reconstruction}. This structural relaxation plays a crucial role in determining the system’s electronic properties by shaping flat bands, modifying band topology, and enhancing electron-phonon interactions \cite{nam2017lattice, carr2020electronic, wang2024fractional, zhang2024polarization}. Dynamically modulating moir\'{e} PLD via photoexcitation could therefore unlock new opportunities to probe and manipulate quantum phases of matter on demand. 

Transition metal dichalcogenide (TMDC) bilayers offer a particularly compelling platform, with strong light-matter interactions and long-lived moir\'{e} excitons \cite{seyler2019signatures,tran2019evidence,jin2019observation,alexeev2019resonantly}.
In TMDC moir\'{e} superlattices, charge carriers are expected to form small polarons similar to a charge-density-wave state, in which lattice motion due to exciton formation enhances the spatial modulation of the exciton wave-function \cite{campbell2022exciton, arsenault2024two, biswas2024exciton, dai2024excitonic}. These intertwined electronic and lattice dynamics provide a promising pathway toward light-driven control of moir\'{e} lattice distortions.

However, despite considerable theoretical progress and growing interest, direct experimental investigation of the photoinduced lattice dynamics in moir\'{e} structures remains limited. Most experimental work has relied on optical probes, including photoluminescence \cite{jin2019observation}, transient absorption and transient reflection spectroscopy \cite{barreOpticaldoi:10.1126/science.abm8511, arsenault2024two}, which are not directly sensitive to lattice motion. In contrast, ultrafast electron diffraction (UED) is the ideal direct experimental tool for precise characterization of the transient lattice structural dynamics. However, the application of UED to twisted bilayer moir\'{e} structures has been held back by significant challenges due to the small diffraction volume, the microscopic size of the high-quality twisted bilayers, and the stringent brightness requirements for the pulsed electron probe necessary to resolve fine details in the momentum space of the moir\'{e} mini-Brillouin zone. 

In this work, we perform UED measurements on twisted WSe$_2$/MoSe$_2$ moir\'{e} structures, taking advantage of deterministic large-area TMDC monolayer exfoliation and assembly \cite{liu2020disassembling}, together with a unique electron beam source, collimation and lensing design \cite{duncan2023multi}. We identify clear moir\'{e} electron diffraction features: satellite peaks dressing the main Bragg peaks. The time-dependent modulation of the satellite diffraction peaks offers precise evidence of the evolution of the moir\'{e} PLD upon photoexcitation. The results demonstrate that the local twist angle dynamically evolves as the interlayer separation changes upon carrier injection, consistent with the prediction that twist angle and interlayer separation are intertwined \cite{qiu2024atomic}. We also directly identify a photoinduced tranverse and coherent sub-THz torsional motion with the same spatial periodicity as the moir\'{e} superlattice, in agreement with the predictions of our driven-oscillator model.  The results open the door to optical control of the moir\'{e} structures as a means of manipulating their associated excitonic and strongly-correlated collective electronic behavior \cite{tang2020simulation, arsenault2024two, wang2024chiral}.

\section{Results: observation of photoinduced moir\'{e} superlattice motion}

\begin{figure}[ht!]
\includegraphics[width=1.0\linewidth]{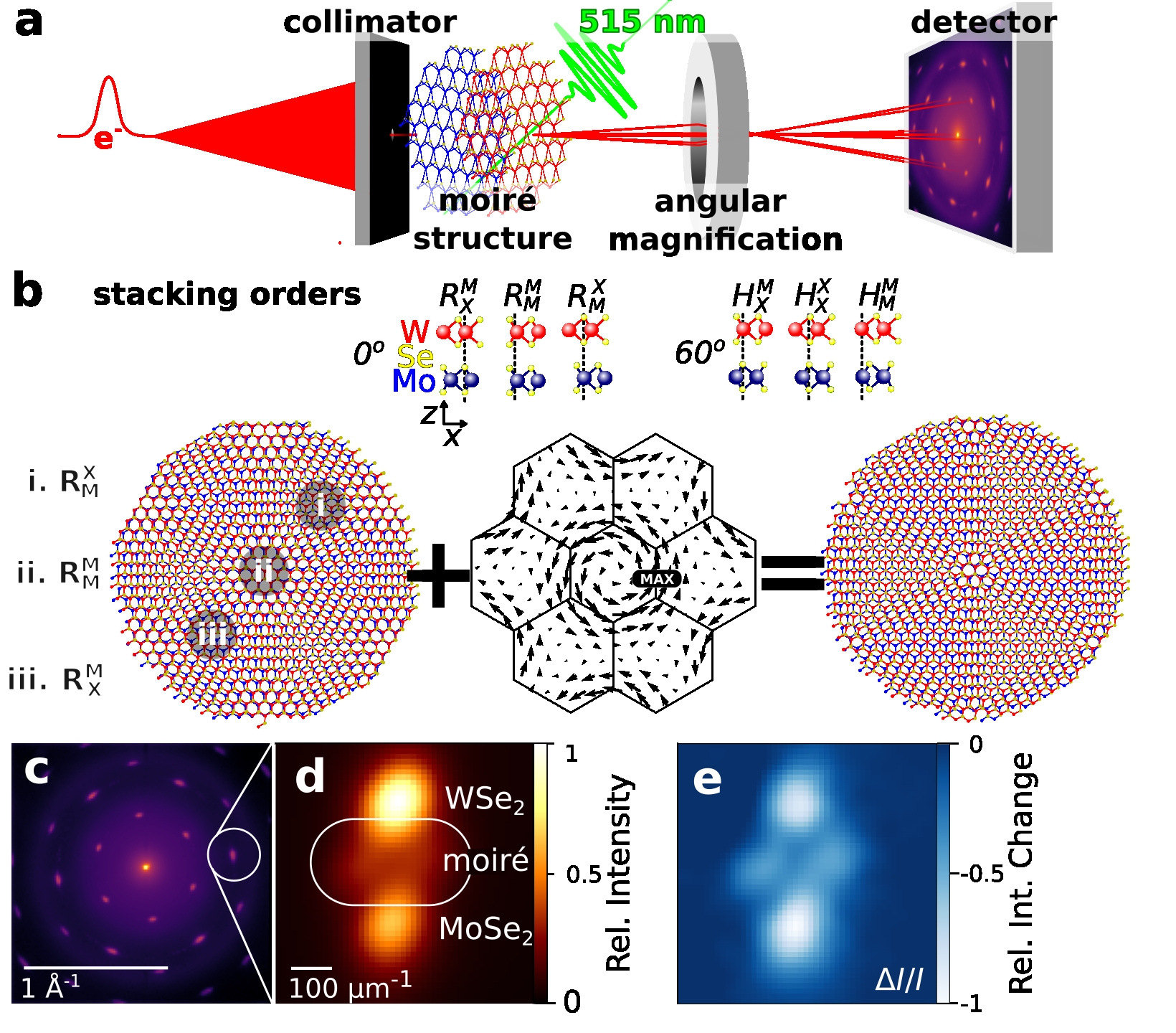}
\caption{\label{fig:diffraction} Experimental scheme and representative diffraction images. (a) Schematics of the UED setup. (b) Periodic lattice distortion (PLD) in a bilayer moir\'{e} structure, consisting of alternating domains with different stacking orders. The PLD expands the size of energetically favorable stacking domains. (c) Wide-angle view of scattering patterns from $2^\circ$ WSe$_2$/MoSe$_2$ heterobilayers. (d) Zoom on the (2,-1) diffraction peak at high angular magnification. (e) Photoinduced changes of scattering intensity, highlighting the strong response of the moir\'{e} satellite peaks. The difference is obtained by subtracting the diffraction pattern 4 ps after photoexcitation from the static pattern in panel (d). The color bar is nolrmalized so that 0 represents no change and -1 the largest decrease in intensity at the delay time shown. Extended Data 1 presents diffraction snapshots at 1 ps for both the $2^\circ$ and $57^\circ$ WSe$_2$/MoSe$_2$ samples.}
\end{figure}

The UED setup is shown schematically in Fig.~\ref{fig:diffraction}(a), and described in detail in the Methods section. Our electron beam source, collimation and lensing system are capable of resolving diffraction features down to 100 $\mathrm{\mu m}^{-1}$ \cite{li2022kiloelectron, duncan2023multi}, effective for resolving the moir\'{e} mini Brillouin zone. The high-quality, large-area WSe$_2$/MoSe$_2$ moir\'{e} structures with twist angles of $2^\circ$ and $57^\circ$ are constructed by stacking macroscopic monolayers from gold tape exfoliation \cite{liu2020disassembling}. Typical diffraction patterns from the large area $2^\circ$ twisted WSe$_2$/MoSe$_2$ moir\'{e} structure are shown in Fig.~\ref{fig:diffraction}(c) and (d). The twist angles are measured via monolayer Bragg peaks in the electron diffraction pattern, and confirmed with second harmonic generation (Extended Data 2). The electron beam is incident normal to the bilayer plane and thus diffraction is only directly sensitive to in-plane motion.

The stacking orders for twisted TMDC bilayers include rhombohedral $R$ type (small twist angles) and hexagonal $H$ type (twist angles close to $60^\circ$). Atomic relaxation leads to an additional local twist \cite{carr2018relaxation}, as illustrated in Fig. ~\ref{fig:diffraction}(b). Consequently, electron diffraction from the moir\'{e} lattice is not merely a superposition of the monolayer Bragg peaks, but exhibits satellite peaks. 

To resolve the moir\'{e} satellite diffraction peaks, we employ projection electron optics to magnify the feature at index (2,-1), shown in Fig.~\ref{fig:diffraction}(c), (d). The magnified view captures both the monolayer Bragg peaks and moir\'{e} satellites. Static electron diffraction experiments performed by others confirm that the moir\'{e}
satellites are due to a torsional PLD, and that the PLD amplitude can be precisely estimated from satellite peak intensities \cite{carr2018relaxation, rosenberger2020twist, sung2022torsional}. The kinematic diffraction theory grounding this well-established result is presented in the Supplementary Information.

Upon excitation with a 515 nm pump laser, the satellite peaks exhibit a more pronounced response relative to the main Bragg peaks, significantly enhancing the contrast, as shown in Fig.~\ref{fig:diffraction}(e). The picosecond time-domain changes of the diffraction pattern for the $2^\circ$ WSe$_2$/MoSe$_2$ and $57^\circ$ WSe$_2$/MoSe$_2$ samples are shown in Fig.~\ref{fig:wiggle}(a)(b)(d)(e). The intensity of the main Bragg peaks decreases upon photoexcitation. Typically, photoexcitation amplifies thermal motion of the lattice, which leads to a reduction in diffraction intensity due to the Debye-Waller (DW) effect \cite{britt2022direct,sood2022bidirectional, johnson2024hidden}. We also observe this behavior in monolayer control samples of WSe$_2$ and MoSe$_2$, shown in Fig.~\ref{fig:wiggle}(g), (h). 

Remarkably, unlike the decay in the main Bragg peaks, the moir\'{e} satellite intensities undergo a pronounced transient increase in the first picosecond after photoexcitation, as shown in Fig.~\ref{fig:wiggle}(a) and (d). This behavior deviates markedly from a purely thermal response, as can be seen from the dashed lines that represent the expected exponential decay from lattice heating alone. Given that the static moir\'{e} satellite peaks arise from an equilibrium torsional PLD \cite{carr2018relaxation, rosenberger2020twist, sung2022torsional}, the observed enhancement indicates a photoinduced amplification of the torsional PLD amplitude. No additional satellites emerge elsewhere in the moir\'{e} Brillouin zone, confirming that the induced motion is predominantly torsional. This conclusion is further supported by rigorous kinematic diffraction theory, detailed in the Supplementary Information, that relies only on the system’s intrinsic threefold rotational symmetry without invoking a specific dynamical model.

To understand why photoexcitation amplifies the torsional PLD, we develop a model in which atomic motion is driven by an out-of-plane force arising from layer-separated photoexcited carriers \cite{wang2019optical, li2023coherent}. The model treats the lattice as a separate subsystem from the electronic degrees of freedom. The lattice dynamics are solved numerically as a driven system of coupled linear ordinary differential equations (Supplementary Information Eq. (S2)). Normal modes of oscillation are computed from molecular dynamics simulations, as described in Methods and illustrated in Extended Data 6. Electronic degrees of freedom are included in the model as a driving force, described by a three-parameter ansatz (Supplementary Information Eq. (S3) and (S4)), which exponentially relaxes from an initial, uniform, out-of-plane pressure $P_i$ to a final, uniform, out-of-plane pressure $P_f$. The fitted relaxation time $\tau\sim1$ ps agrees with carrier lifetimes estimated from time- and angle- resolved photoemission spectroscopy (trARPES, Extended Data 7).  The model combines coherent and thermal effects by first simulating the deterministic motion of a zero-temperature lattice in response to the driving force, and then multiplying the predicted diffraction intensity with a DW-like decaying envelope.  Under our experimental conditions, UED is sensitive primarily to the in-plane projection of this chiral 3D motion. The resulting fits, shown as solid lines in Figs.~\ref{fig:wiggle}(a) and (d), reproduce the overall temporal evolution of both the Bragg and moiré satellite peaks. The best fits for $P_i$ are $\sim0.3$ GPa for $2^\circ$ and $\sim0.2$ GPa for $57^\circ$ WSe$_2$/MoSe$_2$. Statistical analysis of the goodness of fit is presented in Methods and the Supplementary Information, and summarized in Table~\ref{tab:stats}.

Beyond PLD enhancement, the twisted bilayer exhibits coherent oscillations. The corresponding frequency-domain responses, presented in Fig.~\ref{fig:wiggle} (c) and (f), reveal a predominant mode near 0.5 THz. This mode is absent in monolayers on the same Si$_3$N$_4$ substrate, ruling out the acoustic breathing from the sound-wave-round-trip in the substrates \cite{mannebach2017dynamic}. Instead, the observed frequency matches interlayer shear motion observed in the same samples with low-frequency Raman spectroscopy, shown in Fig.~\ref{fig:wiggle}(i). This interpretation is consistent with transient reflectivity measurements reported on WSe$_2$/MoSe$_2$ bilayers, suggesting a sub-terahertz out-of-plane breathing following above-bandgap excitation \cite{li2023coherent}.

By fitting the UED data with our model, we extract the displacement of W, Mo and Se atoms from their initial positions at different delay times. The corresponding real-space snapshots of the $2^\circ$ twisted bilayers are shown in Fig.~\ref{fig:angles}. The appearance of a vortex in the moir\'{e} supercell reflects the dynamical twisting and untwisting of the moire structure, as the coherent driving force couples to a corkscrew-like lattice motion, through the interplay of elastic and van der Waals interactions within the moir\'{e} superlattice. 
To quantify the magnitude of the twisting motion, we evaluate the change in local twist angle with delay time under the small-angle approximation,
\begin{equation}\label{angle}
\theta_\mathrm{local} = \frac{1}{\lvert\lvert {\bf r} \rvert\rvert^{2}} 
\left( \lvert\lvert \Delta{\bf x}_\mathrm{W} \times {\bf r}\rvert\rvert - \lvert\lvert \Delta{\bf x}_\mathrm{Mo} \times {\bf r}\rvert\rvert\right),
\end{equation}
where $\Delta {\bf x}_X({\bf r})$ is the displacement of atomic species $X$ from its position in the rigidly-rotated lattice, and ${\bf r}$ is the position vector taken from the vortex center. To smooth discrete atomic coordinates, a fifth-order polynomial is fit to the expression in parentheses. The limit ${\bf r} \rightarrow 0$ of Eq.~\eqref{angle} is well-defined and gives the best summary of the local atomic environment (Extended Data 9). The time-evolution of the local twist angle due to photoexcitation is shown in Fig.~\ref{fig:angles}(d), indicating a peak-to-trough angle-change of $0.6^\circ$.

 The real-space reconstruction of the lattice motion reveals that shortly after photoexcitation ($\sim 800$ fs), each moiré supercell develops a vortex centered on the $R^M_M$ domain  (Fig.~\ref{fig:angles} (b)). Compared with the initial relaxed structure (Fig.~3 (a)), this additional twist aligns with the equilibrium PLD direction, reinforcing the existing moir\'{e} distortion. At later times (Fig.~\ref{fig:angles} (c)), the local twist reverses sign, reducing the overall moir\'{e} distortion. This evolution is consistent with our experimental observations: the transient increase in satellite peak intensity reflects the initial reinforcement of the PLD, followed by suppression as the moir\'{e} PLD untwists at later time. A qualitatively similar response is observed in the $57^\circ$ sample, shown in Extended Data 8.   

\section{Discussion: Origin of the twist and untwist of moir\'{e} structures locally}

In TMDC bilayers such as WSe$_2$/MoSe$_2$, photoexcitation causes ultrafast interlayer charge transfer \cite{ zhu2017interfacial, ji2017robust, sood2022bidirectional}, as illustrated in Fig.~\ref{fig:mechanism}(a), (b).
Due to the type-II band alignment, photoexcited electrons move to the MoSe$_2$ layer with lower conduction band minimum (CBM) and holes to the WSe$_2$ layer with higher valence band maximum (VBM), leading to layer-separated charges that are attracted by electrostatic forces. Carrier dynamics are validated by complementary trARPES measurements (Extended Data 7) on a separate 3$^\circ$ twisted MoSe$_2$/WSe$_2$ heterobilayer under similar excitation conditions. The trARPES results demonstrate that K-point electrons decay with a 0.6 ps time constant while holes at the K point exhibit biexponential decay with a fast $\sim 2$ ps component and a slower $> 5$ ps component. The fast decay of K-point electrons reflects the relaxation to the CBM at the Q valley of MoSe$_2$/WSe$_2$ \cite{GillenCalculation}, outside the ARPES field of view, whereas the majority of the hole concentration is located at the K point.

We theoretically estimate the interaction strength between photoexcited, layer-seperated carriers and the atomic lattice with \textit{ab initio} electron-phonon calculations, which show that the dominant coupling is to out-of-plane lattice motion (Supplementary Information Table 1). Out-of-plane strain significantly modulates the electronic band structure (the deformation potential), and the change in bandgap $dE_g$ with interlayer distance $dL$ at carrier density $n$ gives rise to a pressure $P_g = (dE_g/dL)n$ \cite{thomsen1986surface}. The result of our \textit{ab initio} simulations is that $dE_g/dL = 1.1 \ \mathrm{eV/nm}$ for CBM electrons and VBM holes (Extended Data 10(d)). Coulomb interactions between layer-separated electrons and holes generate an additional pressure $P_c = (dE_c/dL)n$, where $E_c$ is the average interaction energy per carrier. As detailed in supplementary information, we estimate  the out of plane Coulomb interaction $E_c$ in WSe$_2$/MoSe$_2$ heterobilayers to be on the order of $\sim 100$ meV, corresponding to an attractive force of $dE_c/dL$ = 0.2 eV/nm. 

Summing the above contributions of deformation-potential and Coulomb interaction gives the combined pressure:
\begin{equation}
P_i = \left(\frac{dE_g}{dL} + \frac{dE_c}{dL}\right) n_i,
\end{equation}
where the initial layer-seperated charge density $n_i \sim 1 \times 10^{14} \ \mathrm{cm}^{-2}$, generated under the moderate pump fluence of 1.5 -- 2.5 mJ/cm$^2$ at 515 nm \cite{wang2019optical}. This yields an estimated initial pressure of $\sim$ 0.2 GPa, in good agreement with the values extracted from our UED data. 

Following initial excitation, cooling and decay of charge carriers efficiently reduces the electron-hole population. At delays between 3-4 ps, the carrier density is expected to approach the Mott threshold,  when electrons and holes begin to bind into interlayer excitons (Fig.~\ref{fig:mechanism}(c)). As the carrier density approaches the moir\'{e} density ($1\times 10^{12} \ \mathrm{cm}^{-2}$ for $2^\circ$ WSe$_2$/MoSe$_2$ and $2.5\times 10^{12} \ \mathrm{cm}^{-2}$ for $57^\circ$ WSe$_2$/MoSe$_2$), excitons become trapped at $R^M_M$ and $H^M_X$ sites of the moir\'{e} supercell \cite{schmitt2022formation,karni2022structure}, resulting in a subtle local pinching force. A further source of pressure arises from the Casimir force, on the order of MPa at $n = 10^{12} \ \mathrm{cm}^{-2}$, which does not rely on interlayer charge transfer and persists for a longer time following photoexcitation because it scales with the square root of the photoexcited charge density \cite{mannebach2017dynamic}. Capturing the lattice responses in the dilute-carrier (low pump fluence) regime remains experimentally challenging due to the signal-to-noise ratio, highlighting the motivation for further UED instrument development.

The equilibration of photoexcited carriers with the lattice leads to thermal-expansion pressure via anharmonic coupling to a rising population of optical phonons, as illustrated in Fig.~\ref{fig:mechanism}(d). Using a typical thermal expansion coefficient for monolayer TMDs ($\sim10^{-5}$)
\cite{li2023coherent}, 
the energy injected by the few mJ/cm$^2$ pump fluence would lead to a temperature rise of several tens of Kelvin, yielding a pressure $P_f$ between 100 MPa and 1 GPa. The evolution of net pressure $P(t\geq0) = P_f + (P_i-P_f)e^{-t/\tau}$ that best fits the UED data is plotted in Fig.~\ref{fig:mechanism}(e).

\section{Outlook}
We observe a photoinduced twist that directly modulates the intensity of moir\'{e} satellite diffraction peaks, by first reinforcing and then disrupting the equilibrium torsional displacement in the moir\'{e} structure \cite{carr2018relaxation, rosenberger2020twist, sung2022torsional}.
Out-of-plane electrostatic attraction drives the twist: out-of-plane electronic pressure reduces the interlayer distance, which increases the interlayer vdW interaction 
 and thus
causes atoms to shift in-plane toward a new mechanical equilibrium \cite{ qiu2024atomic},  via an intrinsically chiral motion \cite{wang2024chiral}.

Modulation of the local twist angle provides a mechanism for dynamically controlling the moir\'{e} potential on ultrafast timescales. When the moir\'{e} period exceeds the $\sim 1$ nm exciton Bohr radius, orbitals are well approximated by eigenstates of a harmonic potential: $V({\bf r}) = \frac{1}{2}m\omega^2\lvert\lvert{\bf r}\rvert\rvert^2$. Here ${\bf r}$ is the in-plane displacement from the potential extremum and $m= 0.35 \ \mathrm{m}_e$ is the quasiparticle mass; the frequency is given by $\omega = \theta \sqrt{V_0/(ma_0^2)}$, with $\theta$ the local twist angle near ${\bf r}=0$, $a_0$ the atomic lattice parameter, and $V_0$ a constant determined numerically in the case of WSe$_2$/MoSe$_2$ to be 0.9 eV \cite{wu2018hubbard}. Because low-energy excitons are spatially confined, their spectra are governed by the local rather than global twist angle. The $\pm10\%$ modulation of $\theta$ shown in Fig.~\ref{fig:angles}(d) thus corresponds to a $\pm 4$ meV shift in the ground-state energy of the moir\'{e} trap (from an equilibrium value of $\frac{1}{2}\hbar\omega = 40$ meV), and a $\pm 8$ meV change in the energy of the first excited state. Beyond the harmonic approximation, previous numerical studies have found that lattice reconstruction dramatically alters the energy difference between the moir\'{e} potential minima and maxima \cite{geng2023displacement}, suggesting that this potential depth is likewise tunable via the local twist angle modulation.

Such meV-scale changes to the zero-point energy of the moir\'{e} trap could transiently switch TMDC moir\'{e} bilayers into or out of the superconducting phase. This hypothesis is supported by a Hubbard-model treatment of the recent experimental observations of superconductivity in the twisted homobilayer WSe$_2$ \cite{xia2025superconductivity}, which estimates the superconducting gap at $3.65^\circ$ to be $2 \ \mathrm{meV}$ \cite{kim2025theory}. While the physics of photodoping a homobilayer are expected to differ significantly from a heterobilayer, the same competition between interlayer vdW forces and intralayer strain likely generates a comparable photoinduced twisting motion. The motion could be photoexcited with minimal lattice heating by tuning the pump laser to an exciton resonance. Our results also have implications for exciton-based optoelectronics. Lasing from moir\'{e}-trapped excitons in 2$^\circ$ twisted WSe$_2$/MoSe$_2$ has been recently demonstrated, where to detune the excitonic resonance by 1 meV via the Zeeman effect required an applied field strength of 7 T \cite{qian2024lasing}. The torsional dynamics observed here can achieve the same meV-scale detuning mechanically, without extreme magnetic fields. Finally, because excitonic lifetimes and diffusion rates depend on the moir\'{e} potential depth, ultrafast control of the local twist-angle offers a new pathway for dynamically tuning quantum emitters and excitonic transport in 2D moir\'e materials. 

Our demonstration of ultrafast moir\'{e} twist modulation opens the door to coherent ultrafast control of chiral-acoustic, excitonic, polaronic and, ultimately, strongly-correlated collective degrees of freedom in twisted TMDCs and other twisted bilayers. The experimental and modeling techniques we demonstrate here are equally applicable to a variety of fascinating quasi-2D systems \cite{tang2020simulation, arsenault2024two,  wang2024chiral}.


\newpage

\newpage

\begin{figure}[ht!]
\includegraphics[width=1.0\linewidth]{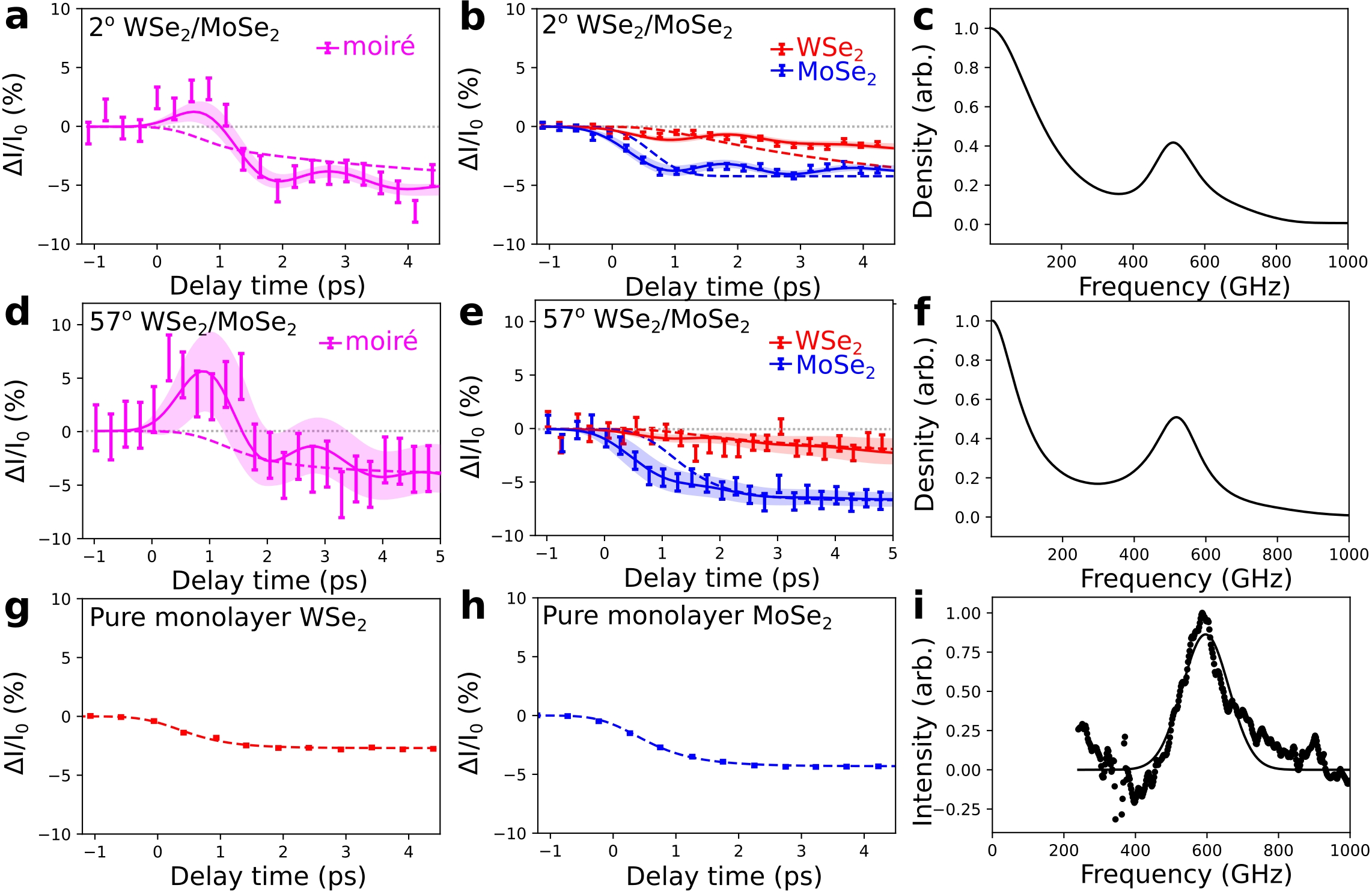}
\caption{\label{fig:wiggle}
Transient changes in diffraction intensity.  Photoexcitation is fixed at 515 nm. Error bars show Poisson uncertainties. Solid lines are fits produced by the dynamical model of Eqs.~(S2)-(S4) in the Supplementary Information, and the shaded area is the 95\% confidence interval. The dashed line is the prediction from lattice heating alone. The diffraction intensity we report is a sum of counts within a detector region of interest that is centered on the relevant peak. (a), (b) Normalized diffraction intensity changes of superlattice peaks and individual monolayer Bragg peaks for $2^\circ$ WSe$_2$/MoSe$_2$ moir\'{e} structure. (d), (e) Normalized intensity changes of $57^\circ $ WSe$_2$/MoSe$_2$ moir\'{e} structure. (g), (h) Normalized intensity changes of isolated WSe$_2$ (g) amd MoSe$_2$ (h) monolayers measured under an identical excitation scheme. (c), (f) Power spectral density of the model for the $2^\circ$ and $57^\circ$ WSe$_2$/MoSe$_2$ moir\'{e} structures (compare discrete Fourier transform of the experimental data, Extended Data 3). (i) Low-frequency Raman spectroscopy of $2^\circ$ WSe$_2$/MoSe$_2$ moir\'{e} structure. Further analysis of the time series data is presented in Extended Data 4, 5.}
\end{figure}

\newpage

\begin{table}
    \centering
    \begin{tabular}{c|c|c|c|c}
          & WSe$_2$ $2^\circ$ & MoSe$_2$ $2^\circ$ & Sat. $2^\circ$ & Mean $2^\circ$ \\
         \hline
        $\chi_1^2 /N$ & 14  & 11 & 4.8 & 9.9 \\
        $\chi_2^2 / N$ & 2.0 & 0.87 & 1.3 & 1.4  \\
        $\chi_3^2 / N $& 2.0 & 0.85 & 1.3 & 1.4 \\
        \hline
          & WSe$_2$ $57^\circ$ & MoSe$_2$ $57^\circ$ & Sat. $57^\circ$ & Mean $57^\circ$\\
         \hline
        $\chi_1^2 /N$ & 1.3 & 2.9 &  1.9 & 2.0 \\
        $\chi_2^2 / N$ & 1.1 & 0.40 &  0.77 & 0.77  \\
        $\chi_3^2 / N $& 1.1 & 0.42 & 0.75 & 0.75
    \end{tabular}
    \caption{Model goodness of fit to UED experimental data. Rows show the sum of squared residuals normalized by empirical variance and number of data points $N$ for three dynamical models and two experimental samples. Columns show the individual contributions from each monolayer Bragg peak, satellite peaks and, in the final column, the sum of all three. We compute an F-test statistic from these tabulated values in the Methods: a heuristic summary is that a table entry near 1 indicates a good fit, significantly greater than 1 that the data is poorly fit and significantly less than 1 that the data is overfit. $\chi_1^2$ is the Debye-Waller-only model (dashed lines in Fig.~\ref{fig:wiggle}(a), (b), (d), (e), (g), (h)). $\chi_2^2$ adds a spatially-uniform driving force (solid lines in Fig.~\ref{fig:wiggle}(a)--(f)). $\chi_3^2$ adds a localized force at the moir\'{e} exciton trapping sites (not shown in Fig.~\ref{fig:wiggle}). More details are provided in the Methods and Supplementary Information.}
    \label{tab:stats}
\end{table}

\newpage

\begin{figure}[ht!]
\centering
\includegraphics[width=1.0\linewidth]{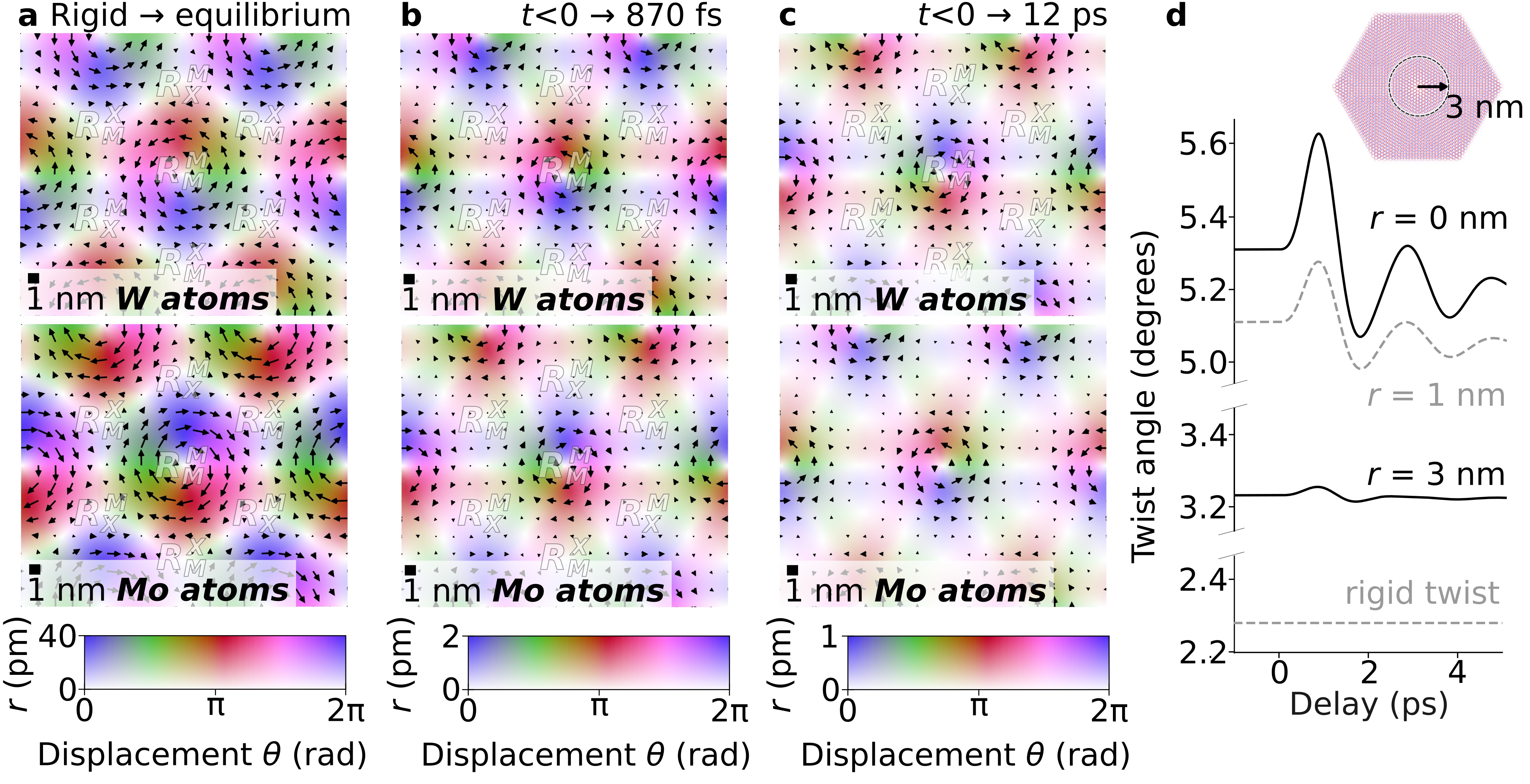}
\caption{\label{fig:angles} Twisting of the $2^\circ$ WSe$_2$/MoSe$_2$ lattice versus time, extracted from our dynamical model fitted to  UED data. (a) At equilibrium, atoms are displaced from sites of the rigidly rotated lattice by vdW forces. Color map indicates local atomic displacement in polar coordinates: hue shows displacement direction $(\theta)$ in radians, saturation displacement magnitude $(r)$ in picometers. (b) Snapshot of the transverse displacement of atoms from equilibrium at 800 fs after photoexcitation, dominated by torsion about the $R_M^M$ stacking domain center. (c) Snapshot at 12 ps following photoexcitation, i.e., after the decay of the oscillatory transient, showing the untwisting of the equilibrium reconstruction. (d) Fitted atomic displacement as a function of radial distance from the vortex center, expressed as a twist angle, defined in Eq.~\eqref{angle}. A similar figure for the $57^\circ$ case is shown in Extended Data 8. A line-out of the displacement modulation versus radial distance is shown in Extended Data 9.}
\end{figure}

\newpage

\begin{figure}[ht!]
\includegraphics[width=1.0\linewidth]{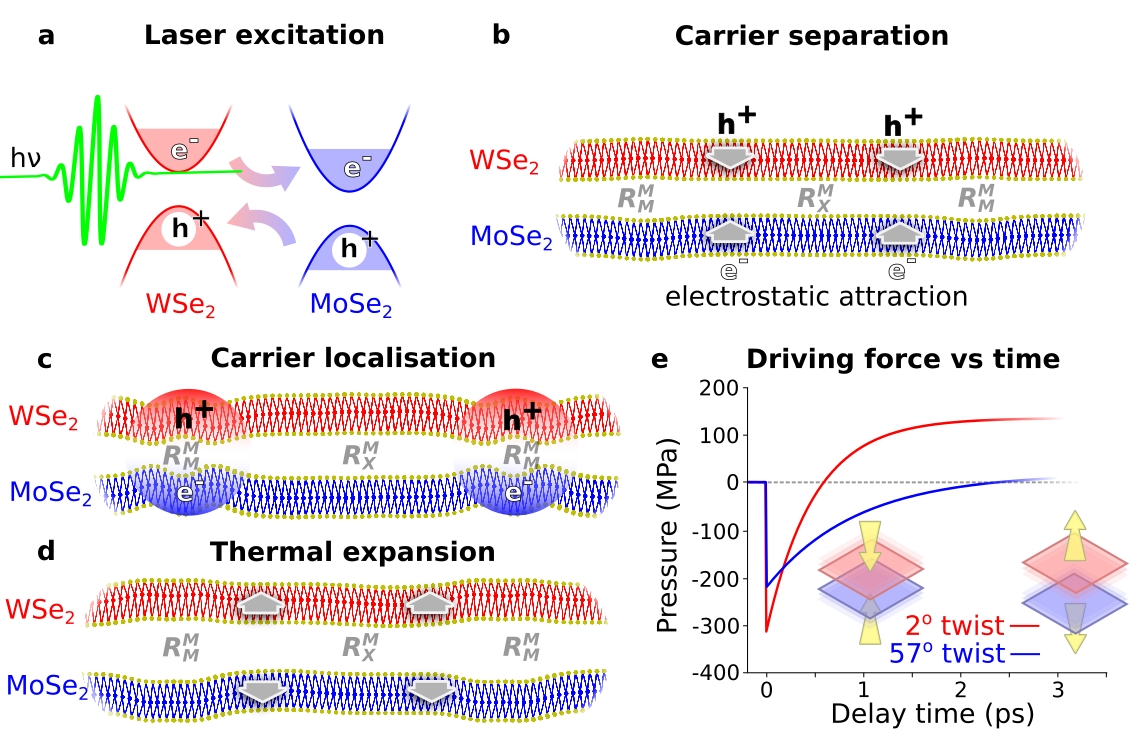}
\caption{\label{fig:mechanism} Charge transfer mechanism driving the lattice response: (a) 515 nm pump photons promote charges into the conduction band; type-II band misalignment causes electrons to move to the MoSe$_2$ layer and holes to WSe$_2$. Bands shown here are schematic: results from density functional theory calculations are presented in Extended Data 10. (b)-(c) Electrostatic forces on layer-separated charges pull the layers together. Simulated atomic displacements from equilibrium are exaggerated in the figure by a factor 100. (d)-(e) Carrier relaxation causes the lattice to heat, and thermal expansion counters the effect of electrostatic attraction. (e) The best-fit driving force that results in the solid curves shown in Fig.~\ref{fig:wiggle}.}
\end{figure}

\newpage
$ $
\newpage
$ $
\newpage
$ $
\newpage
$ $
\section{Methods}

\subsection{UED} 

Probe electron pulses are generated via photoemission from a Na–K–Sb photocathode using 650~nm, 10 ps laser pulses. The probe primary energy is $140 \ \mathrm{keV}$. A radiofrequency cavity compresses the electron pulses to $<$1 ps duration at the sample plane. Sample pumping is performed with $515 \ \mathrm{nm}$, 300 fs pulses. The probe spot size on the sample is 3.5 $\mu$m r.m.s. and the pump spot size 10 $\mu$m r.m.s.

\subsection{Sample preparation}

MoSe$_2$ and WSe$_2$ monolayers are exfoliated from bulk MoSe$_2$ and WSe$_2$ single crystals (HQ graphene)
onto 285~nm $\mathrm{SiO}_2/\mathrm{Si}$ substrate sequentially using a gold tape exfoliation
technique~\cite{liu2020disassembling}, forming
heterostructures with lateral dimensions of mm scale. The crystal orientations of the monolayers in the
heterostructure are aligned with the crystal edges, and further confirmed in electron diffraction and second-harmonic generation. The
heterostructures are later transferred onto 10~nm thick, 250 $\mathrm{\mu m} \times 250 \ \mathrm{\mu m}$ Si$_3$N$_4$ windows on TEM grids
(SiMPore), using a wedging transfer technique with cellulose acetate butyrate
(CAB) polymer, and cleaned with ethyl acetate solvent~[51]

\subsection{Phonon spectra calculations}
We used classical interatomic potentials to describe interactions within individual layers of MoSe$_2$ and WSe$_2$ and between the layers. The intralayer interactions in WSe$_2$ and MoSe$_2$ were described using a Stillinger-Weber potential [52], while the interlayer interactions are modeled using the Kolmogorov-Crespi (KC) potential [10]. We reproduce an expression for the KC potential here:
\begin{equation}
V\left({\bf r}_{ij}, {\bf n}_i, {\bf n}_j\right) = e^{-\lambda(r_{ij}-z_0)}F(\rho_{ij}) - A\left(\frac{r_{ij}}{z_0}\right)^{-6},
\end{equation}
where $r_{ij}$ is the 3D distance between atoms $i,j$, $\rho_{ij}$ is the 2D in-plane projected distance between atoms $i,j$, $z_0$, $\lambda$ and $A$ are free parameters and $F$ a function of the in-plane distances alone.

The relaxed atomic positions of our twisted bilayer systems were determined through the implementation of these potentials within the LAMMPS code [11]. The generation of rigidly twisted heterobilayer structures and subsequent relaxations were automated using the TWISTER package [12]. Phonon frequencies and polarization vectors were obtained by diagonalizing the dynamical matrix with a modified version of the PHONOPY code [13].

\subsection{Goodness of fit}

We fit our UED data with a sequence of three nested models. Here we summarize the detailed quantitative description of the three models that can be found in the Supplementary Information: model 1 considers only the Debye Waller effect (photoinduced reductions in UED intensities due to lattice heating have been reported in other 2D systems including monolayers and bilayers at higher twist angles [35, 36, 37, 57]), model 2 assumes a spatially uniform driving force on the lattice and model 3 introduces a spatial modulation to the driving force. To rank the performance of model $i$, we quantify the lack of fit with a $\chi_i^2$ test statistic:
\begin{equation}
\chi^2_i = \sum^N_{n=1} \left(\frac{\Delta I}{I}(t_n) - \frac{\Delta \hat{I}_i}{\hat{I}_i}(t_n)\right)^2/S^2  
\end{equation}
where $I$ is the measured probe current, $\hat{I}_i$ is the model prediction, $S$ is the empirically estimated r.m.s. uncertainty of the data, and $t_n$ belong to the set of $N$ pump-probe delay times measured. The observed value of these test statistics are shown in Table \ref{tab:stats}.

The null hypothesis is that the residuals are independent and normally distributed with variance $S^2$, which implies that $\chi_i^2/N = 1$. As described in more detail in the Supplementary Information, the significance of the fit improvement can be quantified with the $F_{ij}$ (Fisher) statistic  [22]. We compute the probability that the measured improvement $\bar{F}_{ij}$ (of the fit provided by model $i$ over that of model $j$) is an artifact of experimental uncertainty. For the 2 degree sample, we calculate,
\begin{align}
P\left(F_{12} > \bar{F}_{12}\right) &= 10^{-16}, \\
P\left(F_{23} > \bar{F}_{23}\right) &= 0.8.
\end{align}

For the 57 degree sample, we calculate,
\begin{align}
P\left(F_{12} > \bar{F}_{12}\right) &= 10^{-13}, \\
P\left(F_{23} > \bar{F}_{23}\right) &= 0.2.
\end{align}
These results show that the fit improvement going from model 1 to 2 is unambiguously significant, while the fit improvement from model 2 to 3 is not. Hence, the main text in Figs. 2 and 3 reports the results of model 2.

[51] Schneider, G. F., Calado, V. E., Zandbergen, H., Vandersypen, L. M., Dekker, C.: Wedging transfer of nanostructures. Nano Lett. \textbf{10}(5), 1912--1916 (2010)

[52] Jiang, J.-W., Zhou, Y.-P.: Handbook of Stillinger-Weber Potential Parameters for Two-Dimensional Atomic Crystals. IntechOpen, Rijeka (2017)

[53] Naik, M. H., Maity, I., Maiti, P. K., Jain, M.: Kolmogrov-Crespi Potential for multilayer transition-metal dichalcognides: capturing structural transformations in moir\'{e} superlattices. J. Phys. Chem. C \textbf{123}(15), 9770--9778 (2019)

[54] Thompson, A. P., Aktulga, H. M., Berger, R., Bolintineaunu, D. S., Brown, W. M., Crozier, P. S., In't Veld, P. J., Kohlmeyer, A. Moore, S. G., Nguyen, T. D., Shan, R., Stevens, M. J., Transchida, J. Trott, C. T., Plimpton, S. J.: LAMMPS --- a flexible simulation tool for particle-based materials modeling at the atomic, meso, and continuum scales. Comput. Phys. Commun. \textbf{271}, 108171 (2022)

[55] Naik, S., Naik, M. H., Maity, I., Jain, M.: Twister: Construction and structural relaxation of commensurate moir\'{e} superlattices. Comput. Phys. Commun. \textbf{271}, 108184 (2022)

[56] Togo, A., Tanaka, I: First principles phonon calculations in materials science. Scr. Mater. \textbf{108}, 1--5 (2015)

[57] Hu, J., Xiang, Y., Ferrari, B. M., Scalise, E., Vanacore, G. M.: Indirect exciton phonon dynamics in MoS$_2$ revealed by ultrafast electron diffraction. Adv. Funct. Mater. \textbf{33}(19), 2206395 (2023)

[58] Gonz\'{a}lez-Manteiga, W., Crujeiras, R. M.: An updated review of goodness-of-fit tests for regression models. Test \textbf{22}, 361--411 (2013)

\section{Acknowledgements}
The UED measurements and instrumentation were supported by the U.S Department of Energy, awards DE-SC0020144 and DE-SC0017631, and U.S. National Science Foundation Grant PHY-1549132, the Center for Bright Beams. Preparation of monolayers and twisted heterobilayers at Stanford is supported by the Defense Advanced Research Projects Agency (DARPA) under Agreement No. HR00112390108. F.L. acknowledges support from the U.S. Department of Energy, Office of Science, Basic Energy Sciences, CPIMS Program, under award no. DE-SC0026181. A.M.L. acknowledge support from the U.S. Department of Energy, Office of Science, Basic Energy Sciences, Materials Sciences and Engineering Division, under Contract DE-AC02-76SF00515. The EMPAD detector deployment in this experiment was funded in part by the Kavli Institute at Cornell. The authors are profoundly grateful to Prof. Xiaoyang Zhu for his generous support and guidance in facilitating the trARPES measurements conducted using the setup in his lab at Columbia University. 

C.J.R.D., A.M.L., J.M.M. and F.L. conceived the high-Q-magnification UED experiment. A.M.L., J.M.M. and F.L. supervised the project. C.J.R.D., M.G., A.C.B., and M.K. performed UED measurements. C.J.R.D. and A.C.B. analyzed the UED data to generate the plots shown in main-text Fig.~\ref{fig:wiggle}. C.J.R.D., A.C.B., M.K., W.H.L., M.B.A., C.A.P., I.V.B. and J.M.M. built the UED setup. M.W.T., D.A.M., J.T.L., and S.M.G. built and supported the EMPAD direct electron detector. A.C.J. prepared the samples and performed low frequency Raman measurements with the support of F.L. I.M. simulated the WSe$_2$/MoSe$_2$ phonon spectrum with the support of A.R. All authors discussed the results and contributed to writing the manuscript.

\newpage

\section{Supplementary Information}

\renewcommand{\theequation}{S\arabic{equation}}
\renewcommand{\thetable}{S\arabic{table}}
\let\oldcite\cite
\renewcommand{\citenumfont}[1]{S#1}
\renewcommand{\bibnumfmt}[1]{[S#1]}
\renewcommand{\figurename}{Fig.}
\renewcommand{\thefigure}{S\arabic{figure}}
\setcounter{figure}{0}

\textbf{Contents}

\textbf{1. Band Structure Calculations} \hfill 21

\textbf{2. Linearized Dynamical Model} \hfill 23

\textbf{3. Equivalence of Driving Force with the DECP Formalism} \hfill 26

\textbf{4. Kinematic Diffraction Analysis} \hfill 27

\textbf{5. Goodness of Fit} \hfill 40

\textbf{6. Toy Lagrangian} \hfill 43

\textbf{7. References} \hfill 44

\subsection{Band Structure Calculations}

Extended Data 10 shows the electronic band structure of an untwisted MoSe$_2$/WSe$_2$ bilayer as calculated by density functional theory (DFT). Although the electronic band gap is underestimated at the DFT level, many-body perturbation theory (GW) calculations often lead to a rigid increase in the band gap [S1, S2]. However, \textit{ab initio} GW calculations are computationally expensive and impractical for large systems without major approximations to the screened Coulomb interactions. Previous studies have successfully used DFT to analyze electron wave functions near the band edges of moir\'{e} systems [S3, S4]. Therefore, we employ large-scale \textit{ab initio} DFT calculations to analyze the changes in the electronic structure near the $K$-point band edge for 2.1$^\circ$ and 56.9$^\circ$ twisted MoSe$_2$/WSe$_2$ bilayers. 

We perform the electronic structure calculations using the SIESTA package, which employs localized atomic orbitals as the basis set [S5]. We use norm-conserving Troullier–Martins pseudopotentials [S6], and apply the local density approximation to account for exchange–correlation effects [S7]. We choose a double-$\zeta$ plus polarization basis to expand the wavefunctions. For all moir\'{e} system calculations, we use $\Gamma$-point sampling to obtain the charge density and set a plane-wave energy cutoff of 100 Rydberg. An energy shift of 0.02 Rydberg was applied, along with a vacuum spacing of 20 {\AA} in the out-of-
plane direction. Spin–orbit coupling was included using the onsite approximation [S8].

For the untwisted MoSe$_2$/WSe$_2$ deformation potential calculations, we use a lattice constant of 3.32 {\AA} for
both layers. A finer 15$\times$15$\times$1 k-point grid is used to sample the Brillouin zone, along with a plane-wave
energy cutoff of 350 Rydberg. We displace the atoms following the phonon eigenvector up to 0.04 {\AA}. The vacuum level is subtracted for each calculation before extracting the changes in electronic band energies.

To justify the assumption in our model that the temporally-coherent  
electronic forces on the lattice are purely out-of-plane, we calculate electron-
phonon coupling coefficients for an untwisted WSe$_2$/MoSe$_2$ bilayer with 2H stacking. We compute the change in electronic band structure $D^\nu_{n{\bf K}}$ due to phonons for the valence band and conduction band at the $K$ point of the Brillouin zone:
\begin{equation}
D^\nu_{n{\bf K}} := \frac{\Delta \epsilon_{n{\bf K}}}{su_{\nu \boldsymbol{\Gamma}}}, 
\end{equation}
where $\Delta \epsilon_{n{\bf K}}$ is the change in the electronic band energy at the K point, $u$ is the phonon eigenvector for the $\nu$-th mode at the $\Gamma$ point, and $s$ is a scalling factor. The results are shown in Table~\ref{tab:deformation} for Raman-active modes.

\begin{table}
    \centering
    \begin{tabular}{c|c|c}
         & $D_{\mathrm{C},{\bf K}}^\nu$ (eV/\AA) & $D_{\mathrm{V},{\bf K}}^\nu$ (eV/\AA)\\
         \hline
         Interlayer shear & $-1.6 \times -10^{-4}$ & $-4.8 \times 10^{-3}$ \\
         Interlayer breathing & $-2.8 \times 10^{-2}$ & $1.4 \times 10^{-1}$\\
         ${A^\prime_1}_{\mathrm{MoSe}_2}$ &  2.1 & $-4.9\times10^{-1}$\\
         ${A^\prime_1}_{\mathrm{WSe}_2}$ & $-6.8 \times 10^{-2}$ & $-1.2$\\
         ${E^\prime}_{\mathrm{MoSe}_2}$ & $-8.5 \times 10^{-2}$ & $-4.0\times10^{-3}$\\
         ${E^\prime}_{\mathrm{WSe}_2}$ & $-2.4 \times 10^{-3}$ & $-1.0\times 10^{-1}$
    \end{tabular}
    \caption{\textit{Ab initio} numerically calculated electron-phonon coupling for $0^\circ$ WSe$_2$/MoSe$_2$. The first column labels the phonon mode, the second column shows coupling to the conduction band at the $K$ point, and the final column shows coupling to the valence band at the $K$ point. Optical intralayer modes $A^\prime_1, E^\prime$ carry subscripts indicating the layer affected.}
    \label{tab:deformation}
\end{table}

With respect to the interlayer motions relevant to the moir\'{e}-scale diffraction features, we find that the electron-phonon coupling to the breathing mode is a factor $30$ times stronger than the coupling to interlayer shearing. Table~\ref{tab:deformation} also reports the calculated electron-phonon coupling to optical modes relevant to thermalization of the lattice. These results indicate that the intralayer, out-of-plane $A_1$ modes couple more strongly to free carriers than the intralayer, in-plane $E$ modes, consistent with our interpretation that thermal expansion of the lattice occurs as the free carrier population decays.

\newpage

\subsection{Linearized Dynamical Model}

In this section, we elaborate on further details of the model we use to generate the fits to the data shown in main-text Figs.~2(a)-(h). The model combines coherent and thermal effects by first simulating the deterministic motion of a zero-temperature lattice in response to the driving force, and then multiplying the predicted diffraction intensity with a DW-type exponentially decaying envelope. The simulation treats the displacement of each atom from its initial position in the WSe$_2$/MoSe$_2$ moir\'{e} structures as a driven system of coupled linear ordinary differential equations. Indexing the coordinate axes $(x,y,z)$ as $(0,1,2)$ and writing the $j$th vector component of the displacement of the $i$th atom $x_{ij}$, the system of equations can be expressed:
\begin{equation}\label{model} 
\ddot{x}_{ij} = F_{ij}(t)  -\sum_{k \ell q }U_{kij}U_{k\ell q} \sqrt{\frac{m_{\ell}}{m_{i}}}\left(2\omega_k \dot{x}_{\ell q} +\omega_k^2 x_{\ell q}\right).
\end{equation}
Here, $\omega_i$ represents the normal mode frequencies of lattice oscillation and $m_i$ the mass of the $i$th atom. The tensor $U_{kij}$ is the displacement of the $i$th atom along the $j$th axis when the $k$th normal mode is excited, normalized such that, for all $k$, $\sum_{\ell q} U_{k\ell q}^2 = 1$. The inclusion of a driving term $F_{ij}(t)$ in Eq. \eqref{model} is equivalent to a time-dependent change in the location of the potential energy minima of each atom [9]: a derivation of the equivalent formulation can be found in \S 4 below.

For the driving term $F_{ij}(t)$, we choose the functional form,
\begin{align}
    F_{i0}(t) :=& F_{i1}(t) := 0 \label{noforce},\\
    F_{i2}(t) :=& m_i^{-1}L_i\left[ae^{-t/\tau} + b\left(1-e^{-t/\tau}\right)\right]H(t), \label{force}
\end{align}
where $m_i$ is the mass of the $i$th atom and $L_i$ indicates whether the atom belongs to the WSe$_2$ layer ($L_i = 1$) or the MoSe$_2$ ($L_i = -1$). This functional form can be thought of as the leading term in a more general Fourier expansion of the spatial variation in excited carrier density and electron-phonon coupling across the moir\'{e} supercell. The leading term must be uniform and out-of-plane because of the $c_3$ rotational symmetry of the lattice [S9], which is not broken when free carriers are photoexcited. The factor $L_i$ ensures that the layers are driven in opposite directions.  $H$ is the Heaviside step function, and $a, b, \tau$ are the fit parameters describing the initial ($a$) and final ($b$) strength of the force, and the relaxation time ($\tau$). The model assumes that at acoustic frequencies the lattice is critically damped due to irreversible transfer of energy to the Si$_3$N$_4$ substrate. Normal modes of oscillation are found from molecular dynamics simulations, with potentials fit to agree with density-functional-theory calculations in the neighborhood of each atom [S10, S11, S12, S13].

To derive Eqs~\eqref{model}--\eqref{force}, we start with the time-independent terms on the right-hand-side of Eq.~\eqref{model}, and define a stiffness tensor in terms of the non-perturbative Lagrangian $L$,
\begin{equation}
C_{ijk\ell} := \frac{\partial^2 L}{\partial x_{ij}\partial x_{k\ell}}.
\end{equation}
Then a linearized lattice Lagragian $L_0$ can be defined near equilibrium,
\begin{equation}
L_0 = \frac{1}{2}\sum_{ij} m_j \dot{x}^2_{ij} - \frac{1}{2}\sum_{ijk\ell} C_{ijk\ell}x_{ij}x_{k\ell}.
\end{equation}
The indices $ij$ can be flattened to a single index $I$ over all the real-valued degrees of freedom in the problem. Making the change of variables to mass-weighted degrees of freedom $y_I = x_I\sqrt{\frac{m_I}{\langle m \rangle}}$: 
\begin{equation}
L_0 = \langle m \rangle \frac{1}{2}\sum_{I} \dot{y}_I^2 - \frac{1}{2}\langle m \rangle \sum_{IJ}\frac{C_{IJ}}{\sqrt{m_I m_J}}y_I y_J.
\end{equation}
From the properties of the partial derivative, the matrix $A_{IJ} := C_{IJ}/\sqrt{m_Im_J}$ is real and symmetric and can therefore be diagonalized with eigenvalues $c_i$ by the orthogonal change of basis matrix $U_{iI}$. Transformed degrees of freedom $\eta_i$ (normal mode amplitudes) can then be defined:
\begin{equation}
\eta_i = \sum_{I}U_{iI}y_I.
\end{equation}
The system decouples in these degrees of freedom:
\begin{equation}
L_0 = \frac{1}{2}\langle m \rangle \sum_i  \dot{\eta}^2 - \frac{1}{2}\langle m\rangle \sum_i c_i \eta_i^2.
\end{equation}
The Euler-Lagrange equations are thus,
\begin{equation}
\ddot{\eta}_i = - c_i \eta_i,
\end{equation}
with well known harmonic solutions, such that the normal mode frequencies are $\omega_i = \sqrt{c_i}$. Adding a dampling term gives,
\begin{equation}
\ddot{\eta}_i = -2\omega_i\dot{\eta_i} -\omega^2_i \eta_i.
\end{equation}
We can now substitute back the definition of $\eta_i$ in terms of $x_I$ to obtain,
\begin{equation}
\sum_{I}U_{iI} \sqrt{\frac{m_I}{\langle m \rangle}}\ddot{x}_I = - \sum_{I}U_{iI} \sqrt{\frac{m_I}{\langle m \rangle}}\left(2\omega_i \dot{x}_{I} +\omega_i^2 x_I\right).
\end{equation}
Multiplying through $U^{-1}$,
\begin{equation}
\ddot{x}_J = -\sum_i\sum_{I}U_{iJ}U_{iI} \sqrt{\frac{m_I}{m_J}}\left(2\omega_i \dot{x}_{I} +\omega_i^2 x_I\right).
\end{equation}
Finally, unravelling the indices $IJ$ and relabelling dummy indices,

\begin{equation}
\ddot{x}_{ij} = -\sum_k\sum_{\ell q }U_{kij}U_{k\ell q} \sqrt{\frac{m_{\ell}}{m_{i}}}\left(2\omega_k \dot{x}_{\ell q} +\omega_k^2 x_{\ell q}\right).
\end{equation}
The physical interpretation of the driving term $F_{ij}(t)$ in Eq.\eqref{model} is presented in the Discussion section of the main text. Expanding on the argument for the estimate of Coulomb interactions between layer-separated electrons and holes, the average interaction energy per carrier $E_c$ can be expressed in terms of a polarization surface density ${\bf P}$ and an interaction-mediating electric field ${\bf E}$: $E_c = {\bf P}\cdot{\bf E}/n$. Here ${\bf E}$ denotes the field felt by the average excited carrier due to the sum of interactions with \textit{all} other excited carriers. The dependence of $E_c$ on the interlayer distance $L$ then follows from  ${\bf P} =eLn\hat{\bf z}$, with $n$ the layer-seperated carrier density and $\hat{\bf z}$ pointing out of the plane. Thus $dE_c/dL = E_c/L$. Previous photoluminescence results reveal the dependence of $E_c$ on $n$: as the carrier density is driven to saturation, the average energy released by interlayer recombination remains constant to within a few meV [S14]. The absence of an energy shift suggests that $E_c$ can be extrapolated from the exciton binding energy $E_b$ at low excitation density. Estimates of
$E_b$ in WSe$_2$/MoSe$_2$ vary from $30$ to $300$ meV [S15, S16]: taking a midrange value $E_b = 100$ meV [S17], we obtain $dE_c/dL$ = 0.2 eV/nm. 
\newpage
\subsection{Equivalence of Driving Force with the DECP Formalism}

According to the theory of the displacive excitation of coherent phonons (DECP) [S9], ultrafast lattice motion is caused by a change to the equilibrium coordinates $\bar{x}_{ij}(t)$, where $i$ indexes the atom and $j$ the vector component of the displacement from the non-excited equilibrium. In this formalism, Eq.~\eqref{model} can be written,
\begin{equation}\label{model2} 
\ddot{x}_{ij} = -\sum_{k \ell q }U_{kij}U_{k\ell q} \sqrt{\frac{m_{\ell}}{m_{i}}}\left(2\omega_k \dot{x}_{\ell q} +\omega_k^2 (x_{\ell q} - \bar{x}_{\ell q}(t))\right).
\end{equation}
The equivalence of Eq.~\eqref{model2} to main-text Eq. (1) follows on making the substitution:
\begin{equation}
\bar{x}_{\ell q}(t):=\sum_{rst}U_{r\ell q} U_{rst}\sqrt{\frac{m_{s}}{m_{\ell}}} \omega_r^{-2} F_{st}(t),
\end{equation}
with $F_{ij}(t)$ the driving term  appearing in Eq.~\eqref{model}. Equation \eqref{model2} then becomes,
\begin{equation}
\ddot{x}_{ij} = -\sum_{k \ell q }U_{kij}U_{k\ell q} \sqrt{\frac{m_{\ell}}{m_{i}}}\left(2\omega_k \dot{x}_{\ell q} +\omega_k^2 \left(x_{\ell q} - \sum_{rst}U_{rst}U_{r\ell q} \sqrt{\frac{m_{s}}{m_{\ell}}} \omega_r^{-2} F_{st}(t)\right)\right),
\end{equation}
which after expanding the brackets includes a term,
\begin{align}
\sum_{k \ell q rst}U_{kij}U_{k\ell q} U_{r\ell q} U_{rst} \sqrt{\frac{m_s}{m_i}} \frac{\omega_k^2}{\omega_r^2} F_{st}(t) \notag \\
= \sum_{k rst}U_{kij} \delta_{kr} U_{rst} \sqrt{\frac{m_s}{m_i}} \frac{\omega_k^2}{\omega_r^2} F_{st}(t) \notag \\
= \sum_{kst}U_{kij} U_{kst} \sqrt{\frac{m_s}{m_i}} F_{st}(t) \notag \\
= \sum_{st} \delta_{is}\delta_{jt} \sqrt{\frac{m_s}{m_i}} F_{st}(t) \notag \\
= F_{ij}(t),
\end{align}
where $\delta$ is the Kronecker delta symbol. In taking these steps, use is made of the identities that follow from the orthonormality of the normal modes of oscillation, namely:
\begin{equation}
\sum_{\ell q} U_{k\ell q} U_{r\ell q} = \delta_{kr},
\end{equation}
and,
\begin{equation}
\sum_{k} U_{kij} U_{kst} = \delta_{is}\delta_{jt}.
\end{equation}

\newpage
\subsection{Kinematic Diffraction Analysis}
\begin{figure}[ht!]
\centering
\includegraphics[width=0.9\linewidth]{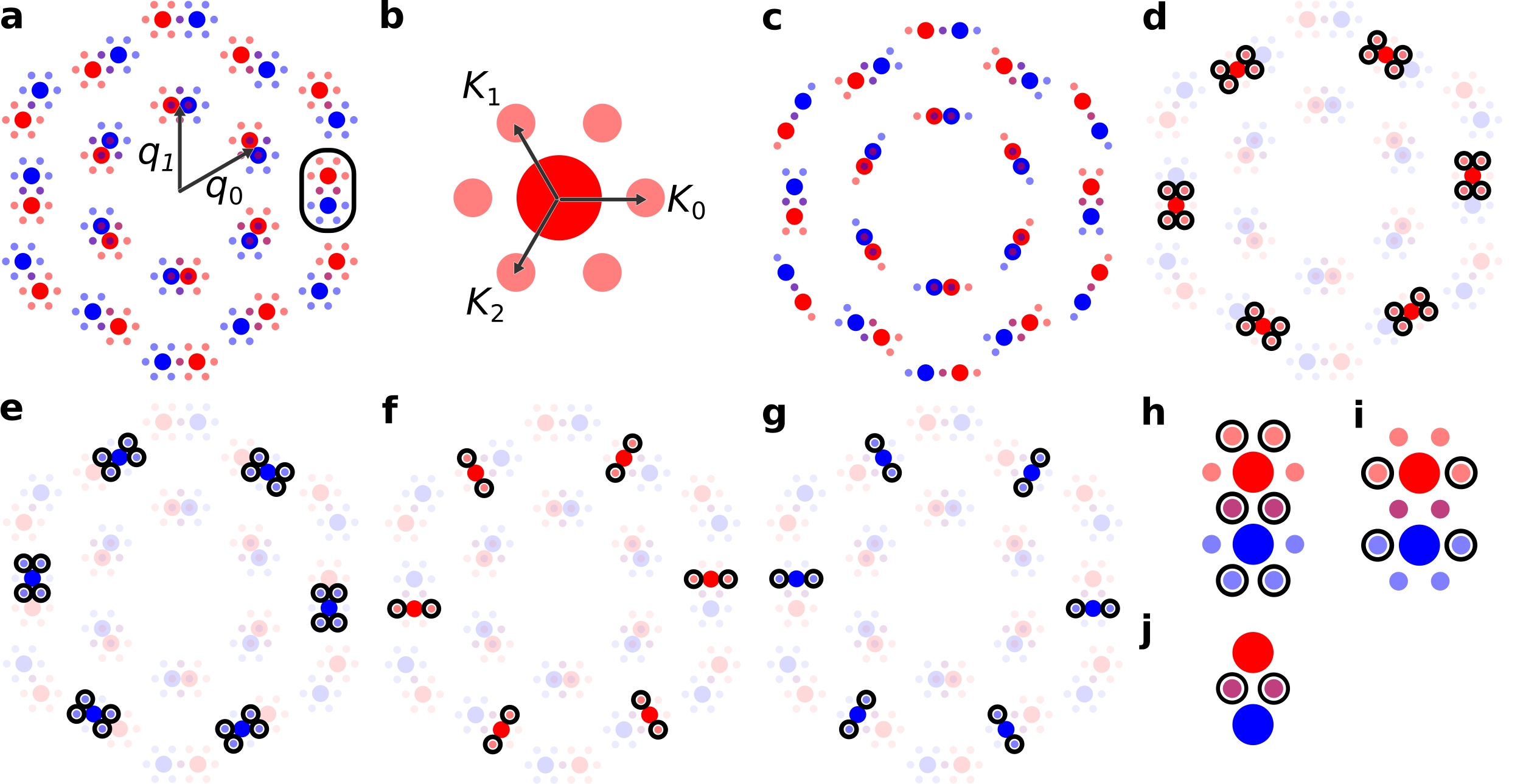}
\caption{(a) Schematic of twisted WSe$_2$/MoSe$_2$ in reciprocal space, for general twist angle, showing all possible first-order moir\'{e} satellites. Contributions from WSe$_2$ are shown in red, MoSe$_2$ in blue. Arrows indicate the atomic-scale basis vectors ${\bf q}_0, {\bf q}_1$. Highlighted is the region of interest at $(2,-1)$, where the UED data presented in the main text is collected. (b) Definitions of the moir\'{e} scale reciprocal lattice vectors. (c) Schematic of twisted WSe$_2$/MoSe$_2$, showing only those satellites with significant intensity when the system is in equilibrium, due to strain waves caused by atomic relaxation. (d)-(e) Torsional PLDs have a significant effect on the highlighted peaks, related to the (2,-1) region of interest by rotation. (f)-(g) Radial PLDs have a significant effect on the highlighted peaks, related to the (2,-1) experimental region of interest by rotation. (h) The highlighted peaks in the (2,-1) experimental region of interest are sensitive to torsional PLDs. (i) The highlighted peaks in the (2,-1) experimental region of interest are sensitive to radial PLDs. (j) The highlighted peaks in the (2,-1) are summed in computing the satellite peak intensities shown in Figs.~\ref{2degirreps} and \ref{57degirreps}. \label{rspace_map}}
\end{figure}

This section derives the expected diffraction signal from the moir\'{e} superlattice.
Assume that the probe electron incident on the sample is a plane wave with wave vector ${\bf k}_0$. Letting ${\bf r}$ be the position on the detector measured with respect to the sample, the solution to the time-independent Schroedinger equation in the first Born approximation is,
\begin{equation}
\psi({\bf r}) = e^{i{\bf k}_0\cdot{\bf r}} + \frac{me}{2\pi\hbar^2}\frac{e^{ik_0r}}{r}\hat{V}(k_0\hat{\bf r}-{\bf k}_0),
\end{equation}
were $\hat{V}$ is the Fourier transform of the sample potential. Our experiment represents the special case in which ${\bf k}_0$ is normal to the bilayer and ${\bf q} := k_0\hat{\bf r} - {\bf k}_0$ is parallel. The potential $\hat{V}$ can be decomposed into the sum of six monolayers $\hat{V}_i$, starting with the bottom Se monolayer of MoSe$_2$ at index 0:
\begin{equation}
\hat{V} = \sum_{i=0}^{5} \hat{V}_i.
\end{equation}
The $\hat{V}_i$ in turn can be decomposed into a product of the the Fourier transform of the potential due to an individual atom $\hat{G}_i$ and the Fourier transform of the crystal lattice $\hat{F}_i$, i.e., letting $r_{ij}$ be the position of the $j$th atom belonging to the $i$th monolayer:
\begin{align}
\hat{F}_i({\bf q}) &= \int_{-\infty}^\infty\int_{-\infty}^\infty\int_{-\infty}^\infty \sum_i\sum_j \delta^3({\bf r} - {\bf r}_{ij}) e^{-i{\bf q}\cdot{\bf r}}dxdydz, \label{lattice}\\
\hat{V}_i &= (\hat{G}_i\hat{F}_i)*\hat{\Omega}. \label{convo}
\end{align}
The indicator function $\Omega({\bf r})$ is unity if ${\bf r}$ lies within the finite bounds of the crystal and vanishes otherwise. Neglecting for the moment the displacement due to moir\'{e}  reconstruction, the $\hat{F}_i$ are all equal up to rotation and translation. Let us therefore define a fiducial $\hat{F}$ as follows:
\begin{equation}
\hat{F}({\bf q}) := \sum_{n=-\infty}^\infty\sum_{m=-\infty}^\infty \delta^3\left({\bf q} -n{\bf q}_0-m{\bf q}_1 \right),
\end{equation}
where the reciprocal lattice vectors ${\bf q}_0, {\bf q}_1$ are defined:
\begin{align}
    {\bf q}_0 &= \frac{2\pi}{a}\hat{\bf x} +\frac{2\pi}{\sqrt{3}a}\hat{\bf y}, \\
    {\bf q}_1 &= \frac{4\pi}{\sqrt{3}a}\hat{\bf y},
\end{align}
and $a$ is the transverse lattice constant for WSe$_{2}$. These basis vectors are illustrated in the context of a schematic diffraction pattern in Fig.~\ref{rspace_map}(a). We make the approximation that the lattice constants of WSe$_2$ and MoSe$_2$ are equal, justified by the fact that the mismatch is negligible compared to the experimental twist angle.
Let $R$ and ${\bf d}$ be respectively the rotation matrix and translation vector,
\begin{equation}
R = \begin{pmatrix} \cos(\theta/2) & -\sin(\theta/2) & \\ \sin(\theta/2) & \cos(\theta/2) & \\ & & 1 \end{pmatrix}, \quad {\bf d} = \begin{pmatrix} a/2 \\ a/\sqrt{12} \end{pmatrix},
\end{equation}
with $\theta$ the rigid moir\'{e} twist angle. We then obtain, when restricted to ${\bf q}$ parallel to the bilayer surface, for the $R$-type stacking:
\begin{align}\label{phasesstart}
    \hat{F}_0({\bf q}) &= e^{i\left(R{\bf q}\right)\cdot {\bf d}} \hat{F}\left(R{\bf q}\right), \\
    \hat{F}_1({\bf q}) &=\hat{F}\left(R{\bf q}\right), \\
        \hat{F}_2({\bf q}) &= e^{i\left(R{\bf q}\right)\cdot {\bf d}} \hat{F}\left(R{\bf q}\right),\\
            \hat{F}_3({\bf q}) &= e^{i\left(R^{-1}{\bf q}\right)\cdot {\bf d}} \hat{F}\left(R^{-1}{\bf q}\right), \\
    \hat{F}_4({\bf q}) &=\hat{F}\left(R^{-1}{\bf q}\right), \\
        \hat{F}_5({\bf q}) &= e^{i\left(R^{-1}{\bf q}\right)\cdot {\bf d}} \hat{F}\left(R^{-1}{\bf q}\right). \label{phasesend}
\end{align}

\begin{figure}[ht!]
\centering
\includegraphics[width=0.9\linewidth]{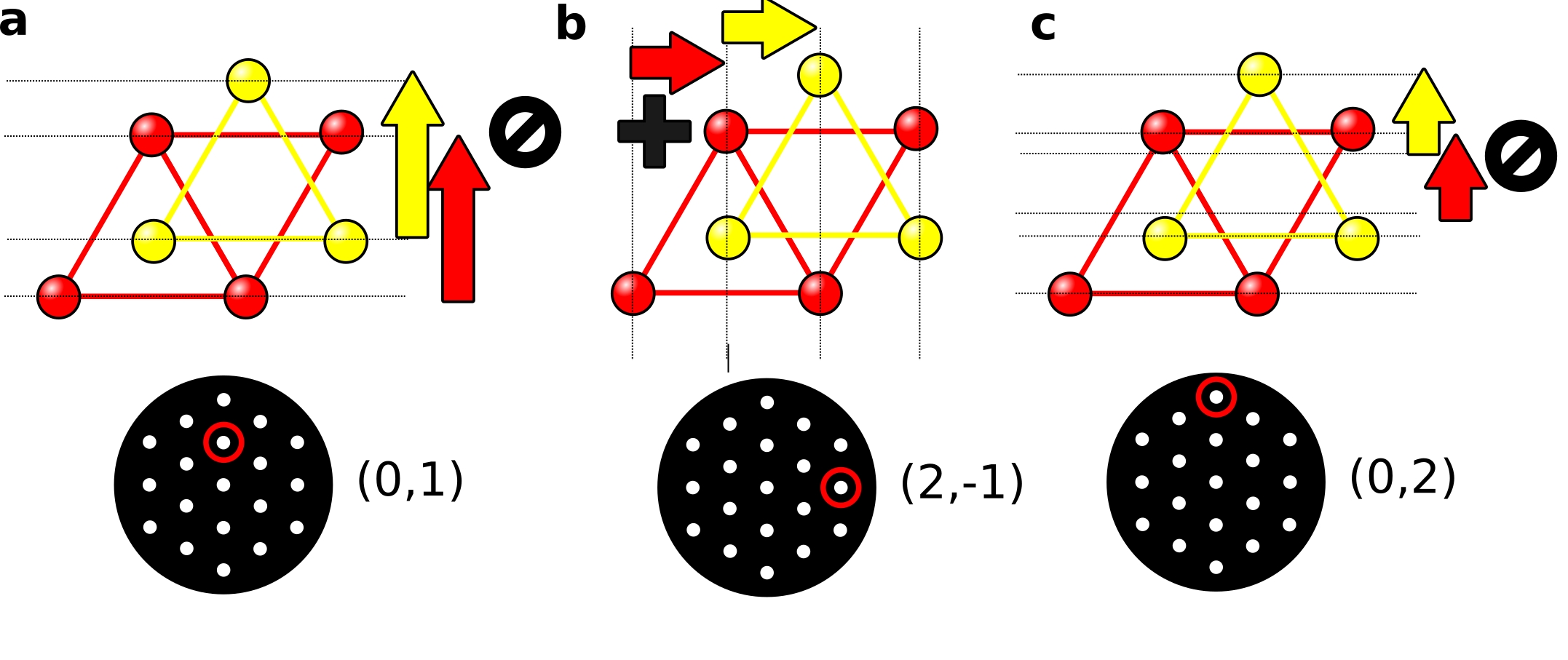}
\caption{Illustration of phase interference due to prefactors in Eqs.~\eqref{phasesstart}--\eqref{phasesend}. (a) Chalcogenide (yellow) and metal (red) imperfectly destructively interfere at diffraction order (0,1). (b) Chalcogenide and metal atoms perfectly constructively interfere at (2,-1). (c) Chalcogenide and metal atoms imperfectly constructively interfere at (0,2). \label{interference}}
\end{figure}

The case of $H$ type stacking is obtained by rotating MoSe$_2$ by $\pi/3$. Although each monolayer is invariant under a $\pi/3$ rotation, the stacked structure is only invariant under a $2\pi/3$ rotation. The effect of a $\pi/3$ rotation is instead equivalent to a permutation, i.e., chalcogenide atoms go to the locations previously occupied by metal atoms and vice versa. Hence, for $H$-type stacking:
\begin{align}
    \hat{F}_0({\bf q}) &= \hat{F}\left(R{\bf q}\right) \\
    \hat{F}_1({\bf q}) &=e^{i\left(R{\bf q}\right)\cdot {\bf d}} \hat{F}\left(R{\bf q}\right) \\
        \hat{F}_2({\bf q}) &=  \hat{F}\left(R{\bf q}\right)\\
            \hat{F}_3({\bf q}) &= e^{i\left(R^{-1}{\bf q}\right)\cdot {\bf d}} \hat{F}\left(R^{-1}{\bf q}\right) \\
    \hat{F}_4({\bf q}) &= \hat{F}\left(R^{-1}{\bf q}\right) \\
        \hat{F}_5({\bf q}) &= e^{i\left(R^{-1}{\bf q}\right)\cdot {\bf d}} \hat{F}\left(R^{-1}{\bf q}\right).
\end{align}

\begin{figure}[ht!]
\centering
\includegraphics[width=0.7\linewidth]{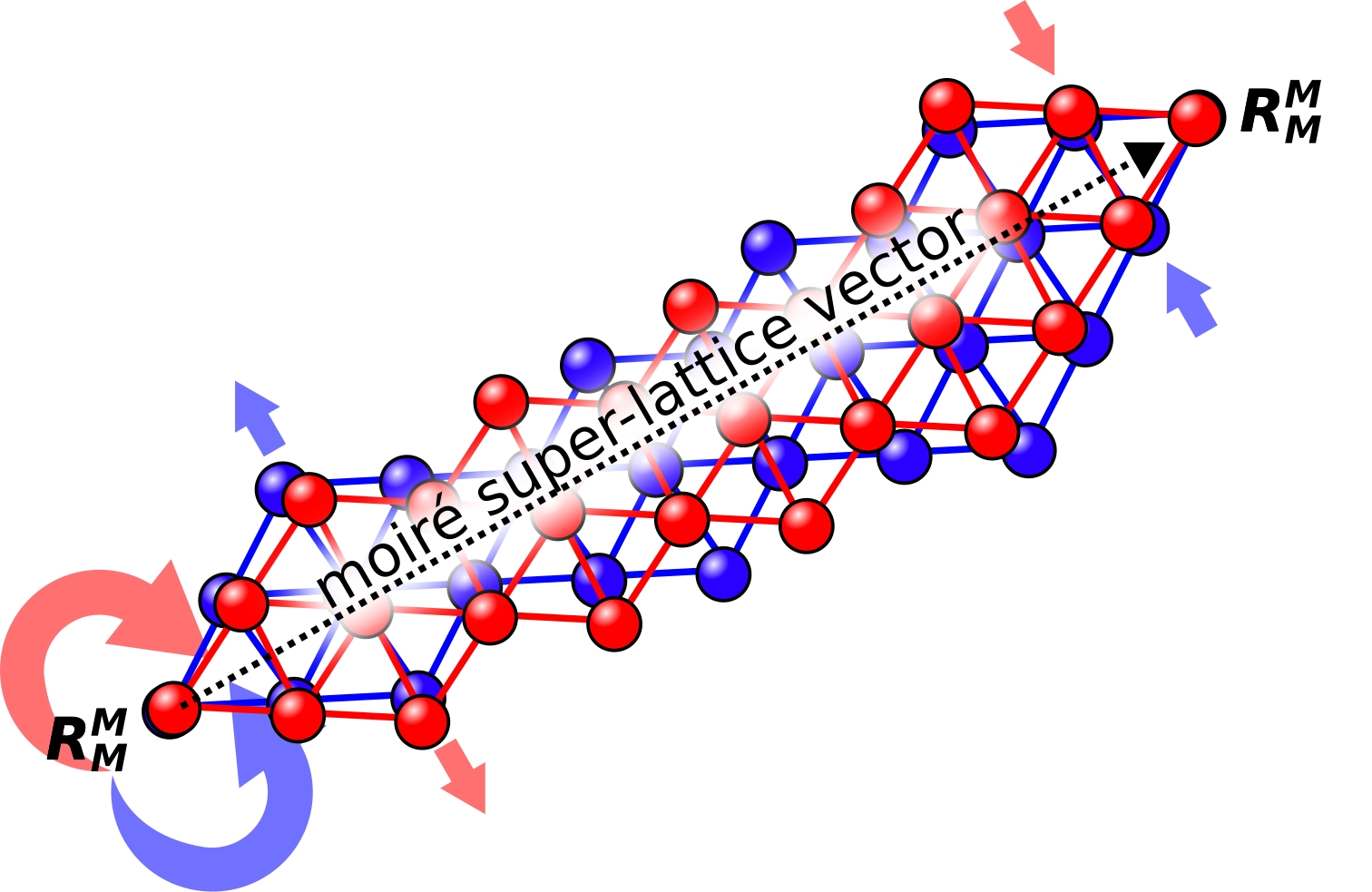}
\caption{Graphical derivation of moir\'{e} superlattice vectors: Eqs.~\eqref{superstart}--\eqref{superend}. In the small angle approximation, a rotation is a motion orthogonal to the vector joining the displaced atom to the axis of rotation. The moir\'{e} supercell can be defined by the repetition of metal-on-metal stacking ($R^M_M$). A metal-on-metal stacking repeats at a distance such that the rotation has displaced an atom by one atomic lattice vector (the shortest path between adjacent atoms). Hence, the vector joining adjacent $R_M^M$ sites is orthogonal to an atomic lattice vector \label{moire_vec}, and the moir\'{e} superlattice is rotated $90^\circ$ relative to the atomic lattice.}
\end{figure}

 In a small angle approximation, near $0^\circ$ in the case of $\theta=2^\circ$ twisted WSe$_2$/MosSe$_2$ and near $60^\circ$ in the case of $\theta=57^\circ$ twisted WSe$_2$/MoSe$_2$, the shift in reciprocal space resulting from the rotation $R$ can be described by three moir\'{e} super-reciprocal-lattice vectors:
\begin{align}\label{superstart}
{\bf K}_0 &:= \frac{4\pi\theta}{\sqrt{3}a}\hat{\bf x}, \\
{\bf K}_1 &:= -\frac{2\pi\theta}{\sqrt{3}a}\hat{\bf x} +\frac{2\pi\theta}{a}\hat{\bf y}, \\
{\bf K}_2 &:= -{\bf K}_0 -{\bf K}_1.\label{superend}
\end{align}
(The third super-reciprocal-lattice vector,
${\bf K}_2$, 
though strictly redundant, will be a helpful notational convenience later.)  These basis vectors are illustrated in the schematic mini-Brillouin zone shown in Fig.~\ref{rspace_map}(b). The derivation of Eq.~\eqref{superstart}--\eqref{superend} is summarized graphically in Fig.~\ref{moire_vec}. Note that the superlattice is locked to be 90$^\circ$ rotated relative to the atomic lattice, which is a consequence of the fact that lattice constants almost exactly match.  In general, we can use the 
shorthand $(i, j, k, \ell)$ to refer to the peak 
at ${\bf q}=i{\bf q}_0 + j{\bf q}_1 + k{\bf K}_0 + \ell{\bf K}_1$. The main WSe$_2$ and MoSe$_2$ Bragg peaks
in the (2,-1) region of interest highlighted in Fig.~\ref{rspace_map}(a) serve as convenient alternate origins for coordinates in reciprocal space, and thus it is helpful to introduce the special notation ${
\bf q}_{
\mathrm{WSe}_2}$ and ${
\bf q}_{
\mathrm{MoSe}_2}$ to refer them:
\begin{align}\label{vecW}
    {\bf q}^\mathrm{ROI}_{
\mathrm{WSe}_2} := 2{\bf q}_0 -{\bf q}_1+ \frac{1}{2}{\bf K}_0 + {\bf K}_1 \\
    {\bf q}^\mathrm{ROI}_{
\mathrm{MoSe}_2} := 2{\bf q}_0 -{\bf q}_1- \frac{1}{2}{\bf K}_0 - {\bf K}_1 \label{vecMo}
\end{align}
(Half-integer indexing of the ${\bf K}_i$ vectors is needed in this notation only because the twist is defined in a symmetrical way, i.e., the layers are rotated by $\pm\theta/2$. Any two peaks in reciprocal space differ by a strictly integer combination of the ${\bf K}_i$, e.g., ${\bf q}^\mathrm{ROI}_{
\mathrm{WSe}_2} - {\bf q}^\mathrm{ROI}_{
\mathrm{MoSe}_2}={\bf K}_0 + 2{\bf K}_1$.  ) 

Directly substituting the reciprocal space coordinates of moir\'{e} satellite peaks into the proceeding expressions for the $\hat{F}_i({\bf q})$ shows that the satellite scattering amplitude is null. To see non-zero diffraction satellites, we must add a moir\'{e} periodic lattice distortion (PLD) to the atomic coordinates. The PLD displacement field ${\bf u}_i({\bf r})$ in layer $i$ can be expanded as a Fourier series in super-reciprocal-lattice vectors ${\bf K}_i$. The satellites visible in our diffraction data all lie a single moir\'{e} period away from a main Bragg peak, and thus only the longest wavelength terms in the Fourier expansion of ${\bf u}_i({\bf r})$ are experimentally relevant. Retaining just these Fourier terms leaves a configuration space with $2\times3\times2=12$ real dimensions, corresponding to the $x$ and $y$ in-plane components of three plane waves in sine and cosine. A yet further reduction in dimensionality can be made because only the subspace of threefold rotationally symmetric ($C_3$ invariant) displacement fields is relevant to our experimental conditions, for the reason that uniform in-plane pump illumination of the probed area preserves the $C_3$ invariance of the equilibrium PLD. To construct all possible $C_3$ invariant displacement fields, we can use group theory to identify the dimensionality of the $C_3$ invariant subspace of the 12 dimensional configuration space, and then proceeded by guessing and checking until we find the required number of basis fields. The 12-dimensional configuration space is large enough to furnish a faithful representation of the dihedral group of order 12 ($D_6$, the symmetries of a hexagon, where ``faithful" means that for every element of $D_6$ there is a unique linear operation on the set of displacement fields). An elementary result of group theory says that there are exactly four one-dimensional irreducible representations of $D_6$ and that these span the $C_3$ invariant subgroup.

\begin{figure}[ht!]
\centering
\includegraphics[width=0.85\linewidth]{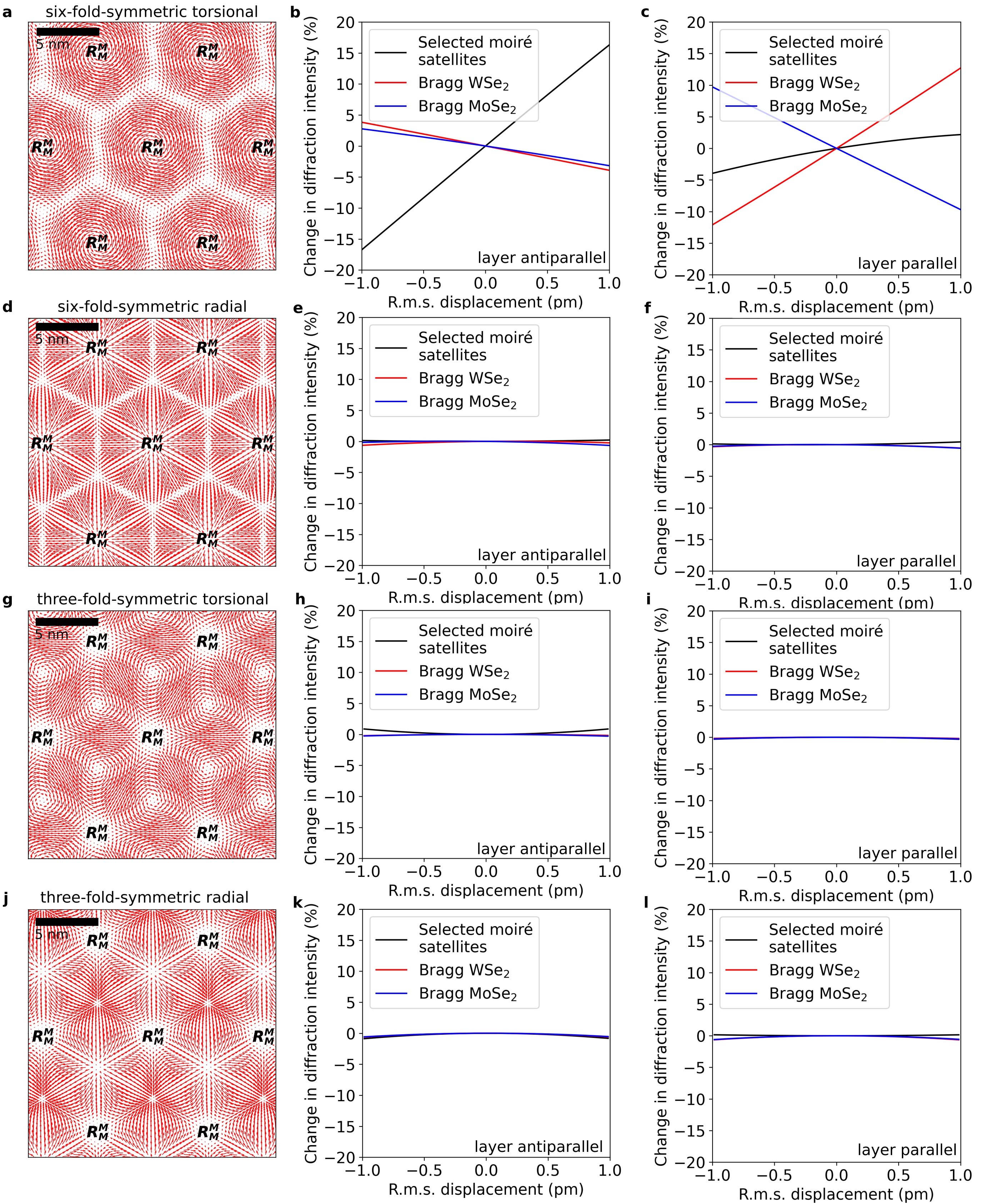}
\caption{\label{2degirreps} Fields spanning all possible threefold symmetric moir\'{e} PLD in $2^\circ$ twisted WSe$_2$/MoSe$_2$.  (a), (d), (g), (j) In-plane visualization of the displacement fields. (b), (e), (h), (k) changes in diffraction intensity with PLD amplitude, added to the static PLD. In red, the WSe$_2$ Bragg peak in the region of interest (ROI) highlighted in Fig.~\ref{rspace_map}(a). In blue, the MoSe$_2$ Bragg peak. In black, the sum of satellite features highlighted in Fig.~\ref{rspace_map}(j). The WSe$_2$ layer moves antiparallel to MoSe$_2$. (c), (f), (i), (l) repeating the same calculations except the WSe$_2$ layer moves parallel to MoSe$_2$.}
\end{figure}

\begin{figure}[ht!]
\centering
\includegraphics[width=0.85\linewidth]{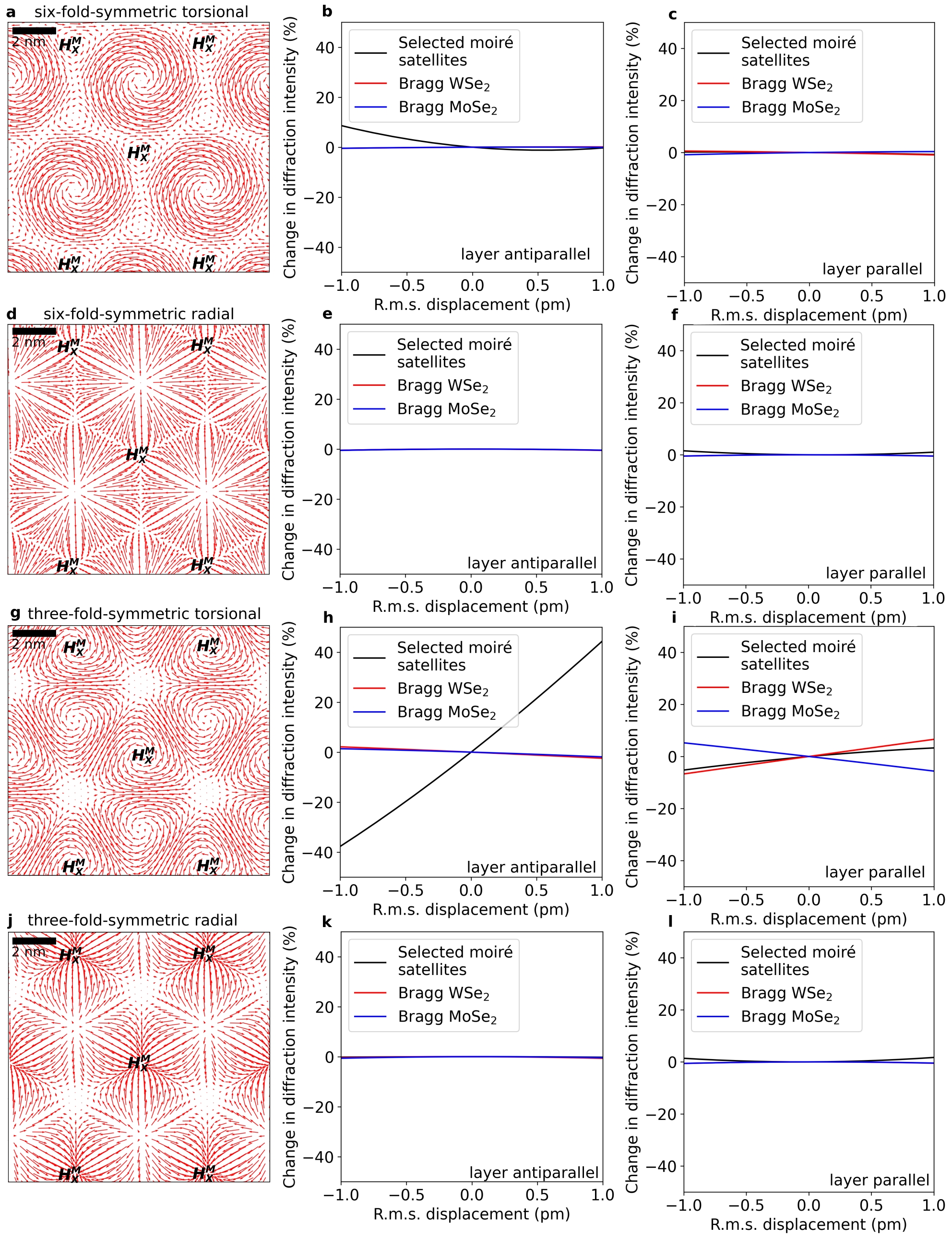}
\caption{\label{57degirreps} Fields spanning all possible threefold symmetric moir\'{e} PLD in $57^\circ$ twisted WSe$_2$/MoSe$_2$.  (a), (d), (g), (j) In-plane visualization of the displacement fields. (b), (e), (h), (k) changes in diffraction intensity with PLD amplitude, added to the static PLD. In red, the WSe$_2$ Bragg peak in the region of interest (ROI) highlighted in Fig.~\ref{rspace_map}(a). In blue, the MoSe$_2$ Bragg peak. In black, the sum of satellite features highlighted in Fig.~\ref{rspace_map}(j). The WSe$_2$ layer moves antiparallel to MoSe$_2$. (c), (f), (i), (l) repeating the same calculations except the WSe$_2$ layer moves parallel to MoSe$_2$.}
\end{figure}

The four displacement fields corresponding to the four one-dimensional irreducible representations are shown in Figs.~\ref{2degirreps} and \ref{57degirreps}  and can be expressed in the following ways. A six-fold rotationally symmmetric torsional motion is described by:
\begin{equation}\label{C6}
{\bf u}^\mathrm{C_6\perp}_i = A_i\nabla \times \sum_{j=0}^2\cos\left({\bf K}_j\cdot{\bf r}\right)\hat{\bf z},
\end{equation}
the $A_i$ parameterizing the amplitude of the displacement. Six-fold rotationally symmetric radial motion is described by: 
\begin{equation}
{\bf u}^\mathrm{C_6\parallel}_i = B_i\nabla \sum_{j=0}^2\cos\left({\bf K}_j\cdot{\bf r}\right),
\end{equation}
with $B_i$ parameterizing the amplitude. Three-fold rotationally symmetric torsional motion is described by:
\begin{equation}\label{C6parallel}
{\bf u}^\mathrm{C_3\perp}_i = C_i\nabla \times \sum_{j=0}^2\sin\left({\bf K}_j\cdot{\bf r}\right)\hat{\bf z},
\end{equation}
parametrized by $C_i$. Finally, three-fold rotationally symmetric radial motion is described by:
\begin{equation}\label{C3}
{\bf u}^\mathrm{C_3\parallel}_i = D_i\nabla\sum_{j=0}^2\sin\left({\bf K}_j\cdot{\bf r}\right),
\end{equation}
parameterized by $D_i$. The symmetry properties of these expressions can be verified by noting that the moir\'{e} reciprocal lattice vectors transform under a rotation by $\pi/3$ as follows:
\begin{equation}
    {\bf K}_0 \mapsto -{\bf K}_2, \quad {\bf K}_1 \mapsto -{\bf K}_0, \quad {\bf K}_2 \mapsto -{\bf K}_1,
\end{equation}
combined with the facts, first, that the expressions above weight all three wavevectors equally and, second, that $\cos(-x)=\cos(x)$ while $\sin(-x)=-\sin(x)$.  The action of the mirror-reflections elements of $D_6$ can be quickly deduced from the properties of the curl and gradient operators: torsional motions are mirror-antisymmetric and radial motions are mirror-symmetric.  Linear combinations of ${\bf u}^\mathrm{C_6\perp}_i, {\bf u}^\mathrm{C_6\parallel}_i, {\bf u}^\mathrm{C_3\perp}_i, {\bf u}^\mathrm{C_3\parallel}_i$ exaust all possible motions, so that there are 24 parameters to be determined from the electron diffraction data, corresponding to $A_i, B_i, C_i, D_i$ for $i \in [0,5]$. The number of degrees of freedom can be reduced to 8 by assuming that,
\begin{equation}
    {\bf u}_0({
\bf r}) \approx {\bf u}_1({
\bf r}) \approx {\bf u}_2({
\bf r}),  \quad {\bf u}_3({
\bf r}) \approx {\bf u}_4({
\bf r}) \approx {\bf u}_5({
\bf r}),
\end{equation}
in other words, atoms belonging to the same TMDC layer are displaced in the same way. This assumption is sound because the intralayer TMDC bonds are much stiffer than the interlayer van der Waals forces.  

It is well established from electron diffraction data and first-principle calculations that the longest wavelength component of the \textit{equilibrium} (i.e. static) PLD in twisted WSe$_2$/MoSe$_2$ at twist angles close to $0^\circ$ is ${\bf u}_i^\mathrm{C_6\perp}$, while for twist angle close to $60^\circ$ the equilibrium PLD is ${\bf u}_i^\mathrm{C_3\perp}$ 
[S18, S19, S20]. 
Figure ~\ref{2degirreps} take the relaxed equilibrium atomic configuration of WSe$_2$ obtained from first-principle calculations (described in the Methods section of the main text) and calculates numerically the change in diffraction intensity that results from modulating the atomic coordinates by adding one of the four displacement fields defined in Eqs.~\eqref{C6}--\eqref{C3}. The intensities plotted are the main Bragg peaks in the experimental region of interest together with the sum of the visible satellites, indicated in Fig.~\ref{rspace_map}. Inspection of the figure shows that  significant change in diffraction intensity occurs if and only if the modulation is the same displacement field as the static PLD. For each of the four basis fields, Fig.~\ref{2degirreps} presents two cases: first, that the TMDC layers move antiparallel and, second, parallel. Panels (b) and (c) show that only antiparallel motion reproduces the feature seen in the experimental data that the satellite peak intensity is \textit{anticorrelated} to the main Bragg peak intensity. Figure \ref{57degirreps} presents the same calculations performed in the $57^\circ$ case. Here, since the static PLD is threefold torsional (due to $H$ type stacking) it is the addition of the threefold torsional PLD that produces the most significant change in diffraction intensity. Again, only antiparallel motion reproduces the anticorrelation between main Bragg and satellite peaks observed in the data.   
\newpage

\newpage
\newpage
Both Figs.~\ref{2degirreps} and \ref{57degirreps} show that the dependence of diffraction intensity on PLD amplitude is close to linear for modulation $\pm 10\%$ of the static diffraction intensity. This linear dependence makes it tractable to explicitly compute expressions for the diffraction intenisty of specific peaks in reciprocal space, which we do below for the $2^\circ$ case. The $57^\circ$ case can be derived by close analogy.

First, we compute scattering due to the six-fold symmetric torsional motion. Expressing the displacement field in terms of the curl of another function makes it simpler to compute the following compact expression for ${\bf q}\cdot {\bf u}_i$, 
\begin{equation} \label{compact}
{\bf q}\cdot{\bf u}_i
= -A_i \sum_{j=0}^2 \left[ {\bf q} \times {\bf K}_j\right]_z\sin\left({\bf K}_j \cdot {\bf r}\right),
\end{equation}
where we use the shorthand notation for the operations on vectors ${\bf v}$, ${\bf w}$,
\begin{equation}
    \left[{\bf v}\times{\bf w}\right]_z := v_xw_y - v_yw_x.
\end{equation}
The local twist angle $\varphi_i$ of layer $i$ around the vortex center can be expressed:
\begin{equation}\label{locald}
    \varphi_i := \lim_{{\bf r}\rightarrow0}\left[\nabla \times {\bf u}_i \right]_z,
\end{equation}
defined so that a counter-clockwise twist is positive.
Applying the vector calculus identity for arbitrary $\bf v$: $\nabla \times \nabla \times{\bf v} \equiv \nabla(\nabla \cdot {\bf v}) - \nabla^2{\bf v}$ to Eq.~\eqref{C6}, and substituting the result into Eq.~\eqref{locald}, yields:
\begin{equation}\label{twistangle}
\varphi_i = 16\pi^2 \frac{\theta^2}{a^2} A_i.
\end{equation}

From Eq.~\eqref{lattice}, and performing the integral,
\begin{equation}\label{step}
    \hat{F}^\prime_i({\bf q}) = \sum_j e^{i{\bf q}\cdot({\bf r}_{ij}+{\bf u}_i)}.
\end{equation}
We now make use of the Jacobi-Anger expansion (a mathematical identity),
\begin{equation}\label{JA}
e^{iA\sin(\phi)} = \sum_{n=-\infty}^\infty J_n(A)e^{in\phi},
\end{equation}
where $J_n$ is the $n$th bessel function.
In the general case of a plane-periodic-strain ${\bf u}_k({\bf r}_{ij})$, with wavevector ${\bf K}_k$ and polarization ${\bf P}_k$, i.e.,
\begin{equation}\label{planewavedef}
    {\bf u}_k({\bf r}_{ij}) = {\bf P}_k\sin\left({\bf K}_k \cdot {\bf r}_{ij}\right),
\end{equation}
Eq.~\ref{JA} reduces to:
\begin{equation}
e^{i{\bf q}\cdot {\bf P}_k \sin({\bf K}_k\cdot{\bf r}_{ij})} = \sum_{n=-\infty}^\infty J_n({\bf q}\cdot {\bf P}_k )e^{in{\bf K}_k\cdot{\bf r}_{ij}},
\end{equation}
so that diffraction sidebands appear at integer multiples of ${\bf K}_k$ with an intensity that is a function of ${\bf q}\cdot {\bf P}_k$. Shifting $\phi \mapsto \phi +\frac{\pi}{4}$ in Eq.~\eqref{JA} results in an analogous formula for cosine,
\begin{equation}\label{cosJA}
e^{iA\cos(\phi)} = \sum_{n=-\infty}^\infty i^nJ_n(A)e^{in\phi}.
\end{equation}
To simplify the analysis we truncated $n\leq 1$ and expand the Bessel function to leading order, so that Eq.~\eqref{JA} reduces to:
\begin{equation}
    e^{iA\sin(\phi)} \approx 1 + \frac{A}{2}\left(e^{i\phi} - e^{-i\phi}\right).
\end{equation}
Substituting this approximation into Eq.~\eqref{step} and retaining terms at most linear in the displacement amplitudes $A_i$,
\begin{equation}\label{almost}
    \hat{F}^\prime_i({\bf q}) = \sum_j\left\{e^{i{\bf q}\cdot{\bf r}_{ij}} - \frac{A_i}{2}\sum_{k=0}^2 
    \left[ {\bf q} \times {\bf K}_k\right]_z 
    \left(e^{i({\bf q}+{\bf K}_k)\cdot{\bf r}_{ij}}-e^{i({\bf q}-{\bf K}_k)\cdot{\bf r}_{ij}}\right)  \right\},
\end{equation}

Recognizing that each of the terms in Eq.~\eqref{almost} is related to the rigid lattice by a translation in reciprocal space ${\bf q} \mapsto {\bf q} \pm {\bf K}_k$, we obtain,
\begin{equation}\label{map}
\hat{F}^\prime_i({\bf q}) = \hat{F}_i({\bf q}) - \frac{A_i}{2}\sum_{k=0}^2  \left[ {\bf q} \times {\bf K}_k\right]_z\left(\hat{F}_i({\bf q}+{\bf K}_k) - \hat{F}_i({\bf q}-{\bf K}_k)\right).
\end{equation}

We now turn to evaluating \eqref{map} at specific points in the moir\'{e} mini-Brillouin zone centered at $(2,-1)$, highlighted in Fig.~\ref{rspace_map}(a). The vectors of the main Bragg peaks satisfy (Eqs.~\eqref{vecW}, \eqref{vecMo}):
\begin{equation}
{\bf q}^\mathrm{ROI}_{
\mathrm{WSe}_2} = \frac{4\pi}{a}R\hat{\bf x}, \quad {\bf q}^\mathrm{ROI}_{
\mathrm{MoSe}_2} = \frac{4\pi}{a}R^{-1}\hat{\bf x}.
\end{equation}
The phase prefactors Eqs~\eqref{phasesstart}--\eqref{phasesend} arising from the displacement between metal and chalcogenide layers are thus,
\begin{equation}
\exp\left\{i(R {\bf q}^\mathrm{ROI}_{
\mathrm{WSe}_2})\cdot {\bf d}\right\} = \exp\left\{i(R^{-1} {\bf q}^\mathrm{ROI}_{
\mathrm{MoSe}_2})\cdot {\bf d}\right\} = \exp\left\{i\left(\frac{4\pi}{a}\hat{
\bf x}
\right)\cdot 
\left(\frac{a}{2}\hat{\bf x} + \frac{a}{\sqrt{12}}\hat{\bf y}\right)\right\}=1,
\end{equation}
and a graphical derivation of this fact is shown in Fig.~\ref{interference}.
It thus follows that,
\begin{align}
    \hat{V}(
    {\bf q}^\mathrm{ROI}_{
\mathrm{MoSe}_2}
    ) = \left(\hat{G}_0 + \hat{G}_1 + \hat{G}_2 \right){\cal N}, \\
        \hat{V}(
{\bf q}^\mathrm{ROI}_{
\mathrm{WSe}_2}
        ) = \left(\hat{G}_3 + \hat{G}_4 + \hat{G}_5 \right){\cal N},
\end{align}
where ${\cal N}$ is the total number of unit cells contained in the volume of the crystal, and arises from the convolution with the indicator function $\Omega$ in Eq.~\eqref{convo}.  

Beginning with satellites offset from the main WSe$_2$ Bragg peak by ${\bf K}_0$, we have for the scattering potential for the satellite peak normalized to the Bragg peak,
\begin{align}
\frac{\hat{V}\left({\bf q}^\mathrm{ROI}_{
\mathrm{WSe}_2}  + {\bf K}_0\right)}{\hat{V}\left({\bf q}^\mathrm{ROI}_{
\mathrm{WSe}_2} \right)} &= \frac{(A_0\hat{G}_0+A_1\hat{G}_1+A_0\hat{G}_1)}{2(\hat{G}_0+\hat{G}_1+\hat{G}_2)}\left[\left( 2{\bf q}_0 - {\bf q}_1 + \frac{3}{2}{\bf K}_0 + {\bf K}_1\right)\times{\bf K_0}\right]_z \\
&= \frac{(A_0\hat{G}_0+A_1\hat{G}_1+A_0\hat{G}_1)}{2(\hat{G}_0+\hat{G}_1+\hat{G}_2)} [{\bf K}_1\times {\bf K}_0]_z \\
&= \frac{(A_0\hat{G}_0+A_1\hat{G}_1+A_0\hat{G}_1)}{(\hat{G}_0+\hat{G}_1+\hat{G}_2)}\frac{4\pi^2}{\sqrt{3}a^2}\theta^2,
\end{align}
in other words, the factor $[{\bf q}\times{\bf K}_0]_z$ is almost zero because the vector ${\bf q}$ is nearly parallel to ${\bf K}_0$. More generally, at the Bragg orders related to $(2, -1)$ by rotation, there is always a reciprocal superlattice point radially offset from the main Bragg peak, and the signature of torsional motion is that this peak vanishes.

Next consider the satellite peak offset from the main WSe$_2$ peak at $(2,-1)$ by ${\bf K}_1$,
\begin{align}
\frac{\hat{V}\left(2{\bf q}_0 - {\bf q}_1 + \frac{1}{2}{\bf K}_0 + 2{\bf K}_1\right)}{\hat{V}\left(2{\bf q}_0 - {\bf q}_1 + \frac{1}{2}{\bf K}_0 + {\bf K}_1\right)} &= \frac{(A_0\hat{G}_0+A_1\hat{G}_1+A_0\hat{G}_1)}{2(\hat{G}_0+\hat{G}_1+\hat{G}_2)}\left[\left( 2{\bf q}_0 - {\bf q}_1 + \frac{1}{2}{\bf K}_0 + 2{\bf K}_1\right)\times{\bf K_1}\right]_z \\
&= \frac{(A_0\hat{G}_0+A_1\hat{G}_1+A_0\hat{G}_1)}{2(\hat{G}_0+\hat{G}_1+\hat{G}_2)} \left([(2{\bf q}_0 - {\bf q}_1)\times{\bf K}_1]_z +\frac{1}{2}[{\bf K}_0\times {\bf K}_1]_z\right) \\
&= \frac{(A_0\hat{G}_0+A_1\hat{G}_1+A_0\hat{G}_1)}{(\hat{G}_0+\hat{G}_1+\hat{G}_2)}\frac{4\pi^2}{a^2}\left(\theta + \frac{1}{2\sqrt{3}}\theta^2\right).
\end{align}
This case includes a term linear in $\theta$, thus the scattering potential for the satellite offset by ${\bf K}_1$ is more intense than the peak offset by ${\bf K}_0$ by a factor $\theta^{-1}$. Now consider the satellite offset by ${\bf K}_2$,
\begin{align}
\frac{\hat{V}\left(2{\bf q}_0 - {\bf q}_1 - \frac{1}{2}{\bf K}_0 \right)}{\hat{V}\left(2{\bf q}_0 - {\bf q}_1 + \frac{1}{2}{\bf K}_0 + {\bf K}_1\right)} &= \frac{(A_0\hat{G}_0+A_1\hat{G}_1+A_0\hat{G}_1)}{2(\hat{G}_0+\hat{G}_1+\hat{G}_2)}\left[\left( 2{\bf q}_0 - {\bf q}_1 - \frac{1}{2}{\bf K}_0\right)\times{\bf K_2}\right]_z \\
&= \frac{(A_0\hat{G}_0+A_1\hat{G}_1+A_0\hat{G}_1)}{2(\hat{G}_0+\hat{G}_1+\hat{G}_2)} \left([(2{\bf q}_0 - {\bf q}_1)\times{\bf K}_2]_z -\frac{1}{2}[{\bf K}_0\times {\bf K}_2]_z\right), \\
&= -\frac{(A_0\hat{G}_0+A_1\hat{G}_1+A_0\hat{G}_1)}{(\hat{G}_0+\hat{G}_1+\hat{G}_2)}\frac{4\pi^2}{a^2}\left(\theta - \frac{1}{2\sqrt{3}}\theta^2\right).
\end{align}
Hence to linear order in the small angle $\theta$, the scattering potential offset from the main Bragg peak by ${\bf K}_2$ is equal in magnitude and opposite in sign to the scattering potential offset by ${\bf K}_1$.

Contributions from WSe$_2$ and MoSe$_2$ overlap at ${\bf q} = 2{\bf q}_0 - {\bf q}_1 \pm \frac{1}{2}{\bf K_0}$. Repeating the above calculations for MoSe$_2$ yields the result that,
 to linear order in the twist angle $\theta$,
\begin{align}\label{sat}
\hat{V}(2{\bf q}_0 - {\bf q}_1 - \frac{1}{2}{\bf K_0}) &= -\frac{4\pi^2\theta}{a^2}\left(A_0\hat{G}_0 + A_1\hat{G}_1 + A_2\hat{G}_2 - A_3\hat{G}_3 - A_4\hat{G}_4 - A_5\hat{G}_5 \right){\cal N}, \\
\hat{V}(2{\bf q}_0 - {\bf q}_1 + \frac{1}{2}{\bf K_0}) &= \frac{4\pi^2\theta}{a^2}\left(A_0\hat{G}_0 + A_1\hat{G}_1 + A_2\hat{G}_2 - A_3\hat{G}_3 - A_4\hat{G}_4 - A_5\hat{G}_5 \right){\cal N}.
\end{align}

Inspection of Eq.~\eqref{sat} shows that, taking the magnitudes of the $A_i$ as given, $\hat{V}$ is maximum when the amplitudes corresponding to WSe$_2$, i.e., $A_1, A_2, A_3$, are opposite in sign to those corresponding to MoSe$_2$, i.e., $A_4, A_5, A_6$.

Substituting into Eq.~\eqref{sat} the expressions given in Eq.~\ref{twistangle} for the local twist angles $\varphi$ about the vortex center results in,
\begin{align}
\hat{V}(2{\bf q}_0 - {\bf q}_1 - \frac{1}{2}{\bf K_0}) &= -\frac{1}{4\theta} \left(\varphi_0\hat{G}_0 + \varphi_2\hat{G}_1 + \varphi_3\hat{G}_2 - \varphi_4\hat{G}_3 - \varphi_5\hat{G}_4 - \varphi_6\hat{G}_5 \right){\cal N}, \\
\hat{V}(2{\bf q}_0 - {\bf q}_1 + \frac{1}{2}{\bf K_0}) &= \frac{1}{4\theta} \left(\varphi_0\hat{G}_0 + \varphi_1\hat{G}_1 + \varphi_2\hat{G}_2 - \varphi_3\hat{G}_3 - \varphi_4\hat{G}_4 - \varphi_5\hat{G}_5 \right){\cal N}.
\end{align}

We can contrast the diffraction signature of six-fold rotationally symmetric torsional motion to a six fold rotationally symmetric radial motion by repeating the same steps starting from Eq.~\eqref{C6parallel}. Replacing the curl with the gradient, the expression for ${\bf q}\cdot {\bf u}_i$ becomes:
\begin{equation} 
{\bf q}\cdot{\bf u}_i
= -B_i \sum_{j=0}^2 {\bf q} \cdot {\bf K}_j\sin\left({\bf K}_j \cdot {\bf r}\right),
\end{equation}
and making the required changes to Eq.~\eqref{map} yields,
\begin{equation}
\hat{F}^\prime_i({\bf q}) = \hat{F}_i({\bf q}) - \frac{B_i}{2}\sum_{k=0}^2  {\bf q} \cdot {\bf K}_k\left(\hat{F}_i({\bf q}+{\bf K}_k) - \hat{F}_i({\bf q}-{\bf K}_k)\right).
\end{equation}
The scattering potential for the peak offset by ${\bf K}_0$ (normalized by the main Bragg peak) is thus,
\begin{align}
\frac{\hat{V}\left({\bf q}^\mathrm{ROI}_{
\mathrm{WSe}_2}  + {\bf K}_0\right)}{\hat{V}\left({\bf q}^\mathrm{ROI}_{
\mathrm{WSe}_2} \right)} &= \frac{(B_0\hat{G}_0+B_1\hat{G}_1+B_2\hat{G}_2)}{2(\hat{G}_0+\hat{G}_1+\hat{G}_2)}\left[\left( 2{\bf q}_0 - {\bf q}_1 + \frac{3}{2}{\bf K}_0 + {\bf K}_1\right)\cdot{\bf K_0}\right]_z \\
&= \frac{(B_0\hat{G}_0+B_1\hat{G}_1+B_2\hat{G}_2)}{(\hat{G}_0+\hat{G}_1+\hat{G}_2)}\frac{8\pi^2}{\sqrt{3}a^2}\left(\theta + \frac{1}{\sqrt{3}}\theta^2\right).
\end{align}
Hence, the effect of the six-fold radial PLD on the diffraction peak at offset ${\bf K}_0$ is stronger by a factor $2\theta^{-1}$ than the sixfold torsonal. For the peaks offset by ${
\bf K}_{1,2}$ we have:
\begin{align}
\frac{\hat{V}\left({\bf q}^\mathrm{ROI}_{
\mathrm{WSe}_2}  + {\bf K}_1\right)}{\hat{V}\left({\bf q}^\mathrm{ROI}_{
\mathrm{WSe}_2} \right)} &= \frac{(B_0\hat{G}_0+B_1\hat{G}_1+B_2\hat{G}_2)}{2(\hat{G}_0+\hat{G}_1+\hat{G}_2)}\left[\left( 2{\bf q}_0 - {\bf q}_1 + \frac{1}{2}{\bf K}_0 + 2{\bf K}_1\right)\cdot{\bf K_1}\right]_z \\
&= -\frac{(B_0\hat{G}_0+B_1\hat{G}_1+B_2\hat{G}_2)}{(\hat{G}_0+\hat{G}_1+\hat{G}_2)}\frac{4\pi^2}{\sqrt{3}a^2}\left(\theta - \frac{7}{2\sqrt{3}}\theta^2\right),\\
\frac{\hat{V}\left({\bf q}^\mathrm{ROI}_{
\mathrm{WSe}_2}  + {\bf K}_2\right)}{\hat{V}\left({\bf q}^\mathrm{ROI}_{
\mathrm{WSe}_2} \right)} &= \frac{(B_0\hat{G}_0+B_1\hat{G}_1+B_2\hat{G}_2)}{2(\hat{G}_0+\hat{G}_1+\hat{G}_2)}\left[\left( 2{\bf q}_0 - {\bf q}_1 - \frac{1}{2}{\bf K}_0\right)\cdot{\bf K_1}\right]_z, \\
&= -\frac{(B_0\hat{G}_0+B_1\hat{G}_1+B_2\hat{G}_2)}{(\hat{G}_0+\hat{G}_1+\hat{G}_2)}\frac{4\pi^2}{\sqrt{3}a^2}\left(\theta - \frac{1}{2\sqrt{3}}\theta^2\right)
\end{align}
Given that the diffraction intensity $I \propto ||\hat{V}||^2$, it follows that the radial PLD induces a factor three stronger effect on the peak offset by ${\bf K}_0$ compared with ${
\bf K}_{1,2}$. In our pump-probe experiments, the constant absence of the ${\bf K}_0$ peaks (highlighted in Fig.~\ref{rspace_map}(i)) indicates that there is no radial component to the photo-induced motion, disambiguating a spiraling motion from a purely twisting motion. 

Expressions for the peak intensity shift due to the threefold torsional PLD can be obtained from the sixfold expression in Eq.~\eqref{map} by introducing the factor $\pm i$ present in Eq.\eqref{cosJA}:
\begin{equation}
\hat{F}^\prime_i({\bf q}) = \hat{F}_i({\bf q}) + i\frac{C_i}{2}\sum_{k=0}^2  \left[ {\bf q} \times {\bf K}_k\right]_z\left(\hat{F}_i({\bf q}+{\bf K}_k) + \hat{F}_i({\bf q}-{\bf K}_k)\right).
\end{equation}
Likewise, the threefold radial PLD is,
\begin{equation}
\hat{F}^\prime_i({\bf q}) = \hat{F}_i({\bf q}) + i\frac{D_i}{2}\sum_{k=0}^2  {\bf q} \cdot {\bf K}_k\left(\hat{F}_i({\bf q}+{\bf K}_k) + \hat{F}_i({\bf q}-{\bf K}_k)\right).
\end{equation}
Hence, the satellite scattering potential of the threefold PLD differs from the sixfold by a factor $\pm i$. 
This imaginary factor implies that a linear-order change in diffraction intensity is only possible if the dynamic torsion matches the symmetry of the static PLD: the steps for reaching this conclusion are as follows. The total scattering potential $\hat{V}_\mathrm{total}({\bf q})$ can be written:
\begin{equation}
\hat{V}_\mathrm{total}({\bf q}) = \hat{V}_\mathrm{static}({\bf q}) + \epsilon \hat{V}_\mathrm{dynamic}({\bf q}),
\end{equation}
where $\epsilon$ is a small parameter. If the static and dynamic PLD have the same symmetries (i.e. both threefold or both sixfold) then the scattering \textit{intensity} $I$ can be expanded in $\epsilon$:
\begin{equation}
I({\bf q}) \propto ||\hat{V}_\mathrm{total}({\bf q})||^2 = ||\hat{V}_\mathrm{static}({\bf q}) + \epsilon \hat{V}_\mathrm{static}({\bf q})||^2 =  ||\hat{V}_\mathrm{static}({\bf q})||^2(1+2\epsilon + \epsilon^2),
\end{equation}
with a term linear in $\epsilon$. If the symmetries differ (ie., static is sixfold but dynamic threefold, or vice-versa), then same expansion yields:
\begin{equation}
I({\bf q}) \propto ||\hat{V}_\mathrm{total}({\bf q})||^2 = ||\hat{V}_\mathrm{static}({\bf q}) + i\epsilon \hat{V}_\mathrm{static}({\bf q})||^2 =  ||\hat{V}_\mathrm{static}({\bf q})||^2(1+ \epsilon^2),
\end{equation}
and there is no term linear in $\epsilon$.

In summary, the vector components of the moir\'{e} PLD can be inferred by comparing the intensity of peaks at the edges of the moir\'{e} mini-Brillouin zone in the experiment region of interest. Two observations are determinative: first, the tangentially offset satellites (Fig.~\ref{rspace_map}(h)) increase in intensity, and second,  the radially offset peaks (Fig.~\ref{rspace_map}(i)) remain absent. By parametrizing the space of all relevant periodic lattice distortions, we find that only torsional motion can produce these observed effects.
\newpage
\subsection{Goodness of Fit}

This section provides a more detailed discussion of the statistical significance of the fits to the UED data shown in main-text Fig.~2. The experimental observables are the relative change in beam current $\Delta I_i (t)/ I_i$ with $i$ indexing the three reciprocal space regions of interest (ROI) indicated in Fig.~1. These ROI are 1. the WSe$_2$ Bragg peak, 2. the MoSe$_2$ Bragg peak and 3. the sum of the two satellite peaks. We fit these data with a sequence of three \textit{nested} models, nested meaning that if $i>j$ then model $i$ contains all parameters present in model $j$. To avoid ambiguity, we denote the model predictions $\Delta \hat{I}_{ij}$ where the index $i$ denotes the ROI and the index $j$ denotes the nest-level of the model. The common element of all three models is an exponential envelope $D_{ij}(t)$, which is the UED response typical of electron-lattice thermal equilibration, defined for ROIs 1 and 2 to be:
\begin{equation}
D_{ij}(t) := \left(A_{ij}\left[e^{(t_0-t)\nu_{ij}}H(t-t_0) + H(t_0 -t)\right] + 1 - A_{ij}\right)*\frac{e^{-0.5t^2/\sigma^2}}{\sigma\sqrt{2\pi}},
\end{equation}
where $A_{ij}$ is the amplitude of the modulation in ROI $i$, $t_0$ is the arrival time of the laser, $\nu_{ij}$ is a relaxation rate, and $\sigma$ is the r.m.s. time resolution of the instrument. The satellite envelope $D_{3,i}$ is the weighted average of monolayer contributions:
\begin{equation}
D_{3,i} = p_1D_{1,i} + p_2D_{2,i} + p_3\sqrt{D_{1,i}D_{2,i}},
\end{equation}
with $p$ given by the ratio of monolayer structure amplitudes, $F_{\mathrm{MX}_2}$:
\begin{align}
    p_1 &= \frac{F_{\mathrm{WSe}_2}^2}{F_{\mathrm{WSe}_2}^2 + F_{\mathrm{MoSe}_2}^2 + 2F_{\mathrm{WSe}_2}F_{\mathrm{MoSe}_2}}, \\
    p_2 &= \frac{F_{\mathrm{MoSe}_2}^2}{F_{\mathrm{WSe}_2}^2 + F_{\mathrm{MoSe}_2}^2 + 2F_{\mathrm{WSe}_2}F_{\mathrm{MoSe}_2}}, \\
    p_3 &= 1 - p_1 - p_2.
\end{align}

Models 2 and 3 involve an extra factor we denote $\Psi_{ij}$ that cannot be expressed in closed form, and is obtained by first solving \eqref{model} 
numerically and then simulating the diffraction pattern 
in the zero temperature, kinematic limit. We perform 
these steps with a custom Python code. The function $\Psi_{ij}$ is normalized to the static diffraction 
intensity such that $\Psi_{i,j}(t<t_0) := 1$ for all 
ROIs. In model 1, there are three additional fitting 
parameters $a_2, b_2, \tau_2$ that describe the spatially-
uniform driving force, according to the expression in
Eq.~\eqref{force}. The most complex model includes an addition spatially-dependent driving force $F_\mathrm{\ell}$,
\begin{equation}
F_\mathrm{\ell}({\bf r}) = c_3\sum_{n}\exp\left\{-\frac{\lvert\lvert{\bf r} -{\bf x}_n \rvert\rvert^2}{2R^2}\right\},
\end{equation}
where ${\bf x}_n$ are the periodic minima in the moir\'{e} potential, $R$ is the r.m.s. size of the trapped states and the fitting parameter $c_3$ quantifies the strength of the force. The predictions of the three models are then,
\begin{align}
\Delta\hat{I}_{i,1}(t) &= \hat{I}_{i,1} D_{i,1}(t), \\
\Delta\hat{I}_{i,2}(t) &= \hat{I}_{i,2} D_{i,2}(t) \Psi_{i,2}(t; a_2, b_2, \tau_2), \\
\Delta\hat{I}_{i,3}(t) &= \hat{I}_{i,3} D_{i,3}(t) \Psi_{i,3}(t; a_3, b_3, c_3, \tau_3),
\end{align}
where $\hat{I}_{ij}$ is the best fit according to model $j$ of the static beam intensity in ROI $i$, i.e., at times $t<t_0$.
The number of fit parameters $\nu_i$ for the $i$th model are therefore $\nu_1 = 4$, $\nu_2 = 7$ and $\nu_3 = 8$.

To rank the performance of model $i$, we quantify the lack of fit with a $\chi_i^2$ test statistic:
\begin{equation}
\chi^2_i = \sum^N_{n=1} \left(\frac{\Delta I}{I}(t_n) - \frac{\Delta \hat{I}_i}{\hat{I}_i}(t_n)\right)^2/S^2  
\end{equation}
where $S$ is the empirically estimated r.m.s. uncertainty of the data, and $t_n$ belong to the set of $N$ pump-probe delay times measured. In a previous publication, we present data analysis verifying that the dominant contribution to experimental uncertainty is Poisson noise, i.e., uncertainty arising from the fact that there are only a finite number of scattering events recorded [S21].  The observed value of these test statistics are shown in Table \ref{tab:sstats}.

\begin{table}
    \centering
    \begin{tabular}{c|c|c|c|c}
         ROI & WSe$_2$ $2^\circ$ & MoSe$_2$ $2^\circ$ & Sat. $2^\circ$ & Sum $2^\circ$ \\
         \hline
        $\chi_1^2 /N$ & 14  & 11 & 4.8 & 9.9 \\
        $\chi_2^2 / N$ & 2.0 & 0.87 & 1.3 & 1.4 \\
        $\chi_3^2 / N $& 2.0 & 0.85 & 1.3 & 1.4 \\
        \hline
         ROI & WSe$_2$ $57^\circ$ & MoSe$_2$ $57^\circ$ & Sat. $57^\circ$ & Sum $57^\circ$\\
         \hline
        $\chi_1^2 /N$ & 1.3 & 2.9 &  1.9 & 2.0 \\
        $\chi_2^2 / N$ & 1.1 & 0.40 &  0.77 & 0.77 \\
        $\chi_3^2 / N $& 1.1 & 0.42 & 0.75 & 0.75
    \end{tabular}
    \caption{Sum of squared residuals normalized by empirical variance and number of data points $N$ for the three dynamical models and two experimental samples. Columns show the individual contributions from each monolayer Bragg peak, satellite peaks and, in the final column, the sum of all three. A value of $1$ indicates good fit, signficantly greater than 1 that the data is poorly fit and significantly less than 1 that the data is overfit. $\chi_1^2$ is the Debye-Waller-only model, $\chi_2^2$ adds a uniform driving force, $\chi_3^2$ adds a localized force at the moir\'{e} exciton trapping sites.}
    \label{tab:sstats}
\end{table}

The null hypothesis is that the residuals are dependent and normally distributed with variance $S^2$, which implies that $\chi_i^2/N = 1$. 

The nesting of the models guarantees that $\chi_3^2 < \chi_2^2 < \chi_1^2$. The significance of the fit improvement can be quantified with the $F_{ij}$ (Fisher) statistic [S22], defined for $j>i$,
\begin{equation}
    F_{ij} := \frac{\chi^2_i - \chi^2_j}{\chi^2_j}\frac{N-\nu_j}{\nu_j -\nu_i}.
\end{equation}
The larger the value of $F_{ij}$, the greater improvement model $j$ makes over model $i$. If the fit residuals are normally distributed with variance $S^2$ (i.e., both models $i$ and $j$ fit the data well) then the $F$ statistic follows the $F$ probability distribution. We can then compute the probability that the measuremed fit improvement $\bar{F}_{ij}$ is an artifact of experimental uncertainty. For the 2 degree sample, we calculate,
\begin{align}
P\left(F_{12} > \bar{F}_{12}\right) &= 10^{-16}, \\
P\left(F_{23} > \bar{F}_{23}\right) &= 0.8.
\end{align}

For the 57 degree sample, we calculate,
\begin{align}
P\left(F_{12} > \bar{F}_{12}\right) &= 10^{-13}, \\
P\left(F_{23} > \bar{F}_{23}\right) &= 0.2.
\end{align}
These results show that the fit improvement going from model 1 to 2 is unambiguously significant, while the fit improvement from model 2 to 3 is not. Hence, the main text in Figs. 2 and 3 reports the results of model 2.

In the case of the $2^\circ$ WSe$_2$/MoSe$_2$ dataset,
the model's central frequency is $510 \pm 90$ GHz, 
overlapping the secondary peak seen in the discrete Fourier transform (DFT) of the experimental time series at 
$540 \pm 90$ GHz (Extended Data~3(b)). 

The central frequency estimated by the model in the 
$57^\circ$ WSe$_2$/MoSe$_2$ dataset is $510 \pm 90$ whereas the secondary peak in the DFT occurs at 
$830 \pm 80$ GHz (Extended Data~3(c)). Inspection of the simulated spectrum of normal modes shows that there is a mode at $877$ GHz (Extended Data~6(g)), which also twists around the vortex center of the static periodic lattice distortion. In our dynamical model, the out-of-plane driving force excites a spectrum of modes that includes the $877$ GHz mode, with the peak being a mode at 528 GHz (Extended Data~6(f)). The r.m.s. displacement of the 877 GHz contribution is, according to the model's best fit, 50\% as large as that of the 528 GHz mode. The puzzle posed by the raw, discrete fourier transform of the UED data is, therefore, not the presence of a component near 877 GHz but rather the absence of a strong component in the bin $500 \pm 80$ GHz. 
A mechanism by which photodoped carriers efficiently couple to the 877 GHz mode without also exciting the 528 GHz mode must involve a force on the lattice that depends intricately on transverse position inside the moir\'{e} supercell. To capture this level of detail in a physical model requires the introduction of additional fit parameters, thereby overfitting the data, as observed above. Ultimately, resolving the puzzle requires the development of brighter electron sources for UED that can deliver improved signal-to-noise at the reciprocal space resolution required to isolate the satellite peak intensities.

\newpage
\subsection{Toy Lagrangian}

The agreement between our experimental data and the model summarized in Eq.~\eqref{model} depends essentially on the coupling between in-plane and out-of-plane lattice degrees of freedom. To better elucidate the coupling mechanism, it is helpful to refer to a toy model that retains only two mesoscopic degrees of freedom: the twist angle $\theta$ about the vortex center in the moir\'{e} supercell  and the spatially averaged interlayer separation $z$. The quadratic Lagrangian $L$ in these variables is,
\begin{align}\label{simple}
L =& \frac{1}{2}I \dot{\theta}^2 + \frac{1}{2}M\dot{z}^2 - \frac{1}{2}I\omega^2_\parallel(\theta-\theta_0)^2- \frac{1}{2}M\omega_\perp^2(z-z_0)^2 \notag\\
&- \frac{1}{2}V\left(\theta - \theta_\mathrm{max}\right)^2+ V\theta\theta_\mathrm{max}\lambda(z-z_0).
\end{align}
Here $I$ and $M$ are the moment of inertia and the mass per moir\'{e} supercell respsectively, 
$\omega_\parallel$ quantifies the strength of the purely intralayer elastic restoring force,
$\omega_\perp$ is the frequency of the purely out-of-plane breathing motion that would be present at zero-twist angle, and $V$ is the vdW energy that gives rise to the PLD. The intralayer elastic equilibrium of the lattice is at the rigid twist angle $\theta_0$, whereas the interlayer vdW equilibrium is $\theta_\mathrm{max}$.
The only addition we make compared with previously published models is the mixing term parameterized by the length-scale $\lambda$ --- a correction implied by the Kolmogorov-Crespi (KC) intermolecular potential that governs the microscopic lattice dynamics (an expression for the KC potential is provide in the Methods section of the main text) [S18, S20, S23]. At our $2^\circ$ and $57^\circ$ twist angles, the out-of-plane van-der-Waals force acting on $z$ is stronger than the in-plane elastic restoring force acting on $\theta$ , so that when the system is fdecoupled $z$ oscillates at a higher frequency than $\theta$ [S24]. The in-plane/out-of-plane hybridization that results from a mixing term in a two-variable, quadratic Lagrangian is straightforward to compute: the lower frequency hybrid mode is mostly in-plane motion and the higher frequency mode mostly out-of-plane motion.

\newpage
\subsection{References}
[S1] F. A. Rasmusen \& K. S. Thygsen. Computational 2D materials database. \textit{J. Phys. Chem. C}, 119(23):13169--13183, 2015

[S2] S. G. Louie \textit{et al.} Discovering and understanding materials through computation. \textit{Nat. Mater.,} 20(6):728--735, 2021

[S3] H. G, X. Zhang \& G. Lu. Shedding light on moir\'{e} excitons. \textit{Sci. Adv.,} 6(42):5638, 2020

[S4] X. Wang \textit{et al.} Intercell moir\'{e} exciton complexes in electron lattices. \textit{Nat. Mater.,} 22(5):599--604, 2023

[S5] J. M. Soler \textit{et al.} The SIESTA method for \textit{ab initio} order-$N$ materials simulation. \textit{J. Phys. Condens. Matter,} 14(11):2745, 2002

[S6] N. Troullier \& J. L. Martins. Efficient pseudo-potentials for plane-wave calculations. \textit{Phys. Rev. B}, 43(3):1993, 1991

[S7] J. P. Perdew \& A. Zunger. Self-interaction correction to density-functional-approximations for many-electron systems. \textit{Phys. Rev. B}, 23(10):5048, 1981.

[S8] L. Fern\'{a}ndez-Seivane \textit{et al.} On-site approximations for spin-orbit coupling in linear combination of atomic orbitals density functional methods. \textit{J. Pys. Condens. Matter,} 18(34):7999, 2006

[S9] H. J. Zeiger \textit{et al.} Theory for displacive excitation of coherent phonons. \textit{Phys. Rev. B,} 45(2):768, 1992.

[S10] M. H. Naik \textit{et al.} Kolmogorov-Crespi potential for multilayer transistion-metal dichalcogenides. \textit{J. Phys. Chem. C,} 123(15):9770--9778, 2019.

[S11] A. P. Thompson \textit{et al.} LAMMPS. \textit{Comput. Phys. Commun.} 271:108171, 2022

[S12] S. Naik \textit{et al.} Twister. \textit{Comput. Phys. Commun.,} 271:108184, 2022

[S13] Atsushi Togo \& Isao Tanaka. First principles phonon calculations in materials science. \textit{Scr. Mater.,} 108:1--5, 2015.

[S14] J. Wang \textit{et al.} Optical generation of high carrier densities in 2D semiconductor heterobilayers. \textit{Sci. Adv.,} 5(9):0145, 2019.

[S15] E. Torun \textit{et al.} Interlayer and intralayer excitons in MoS$_2$/WS$_2$ and MoSe$_2$/WSe$_2$ heterobilayers. \textit{Phys. Rev. B.,} 97(24):245427, 2018

[S16] J. Xia \textit{et al.} Strong coupling and pressure engineering in WSe$_2$--MoSe$_2$ heterobilayers. \textit{Nat. Phys.,} 17(1):92--98.

[S17] M. Van der Donck and F. M. Peeters. Interlayer excitons in transition metal dichalcogenieds heterostructures. \textit{Phys. Rev. B,} 98(11):115104, 2018

[S18] S. Carr \textit{et al.} Relaxation and domain formation in incommensurate two-dimensional heterostructures. \textit{Phys. Rev. B} 98(22):224102, 2018

[S19] M. Rosenberger \textit{et al.} Twist angle-dependent atomic reconstruction and moir\'{e} patterns in transition metal dichalcogenide heterostructures. \textit{ACS Nano,} 14(4):4550--4558, 2020.

[S20] S. H. Sung \textit{et al.} Torsional periodic lattice distortions and diffraction of twisted 2D materials. \textit{Nat. Commun.,} 13(1):7826, 2022.

[S21] C. J. R. Duncan \textit{et al.} Multi-scale time-resolved electron diffraction. \textit{Ultramicroscopy,} 253:113771, 2023

[S22] W. Gonz\'{a}lez-Manteiga \& R. M. Crujeiras. An updated review of goodness-of-fit tests for regression models. \textit{Test}, 22:361--411, 2013.

[S23] A. N. Kolmogorov \& V. H. Crespi. Registry-dependent interlayer potential for graphitic systems. \textit{Phys. Rev. B} 71(23):235415, 2005

[S24] P. K. Nayak \textit{et al.} Probing evolution of twist-angle-dependent interlayer excitons in MoSe$_2$/WSe$_2$ van der Waals heterostructures. \textit{ACS Nano,} 11(4):4041--4050, 2017.

\newpage
\section{Extended Data}

\renewcommand{\figurename}{Extended Data}
\setcounter{figure}{0}

\begin{figure}[ht!] \centering
\includegraphics[width=0.7\linewidth]{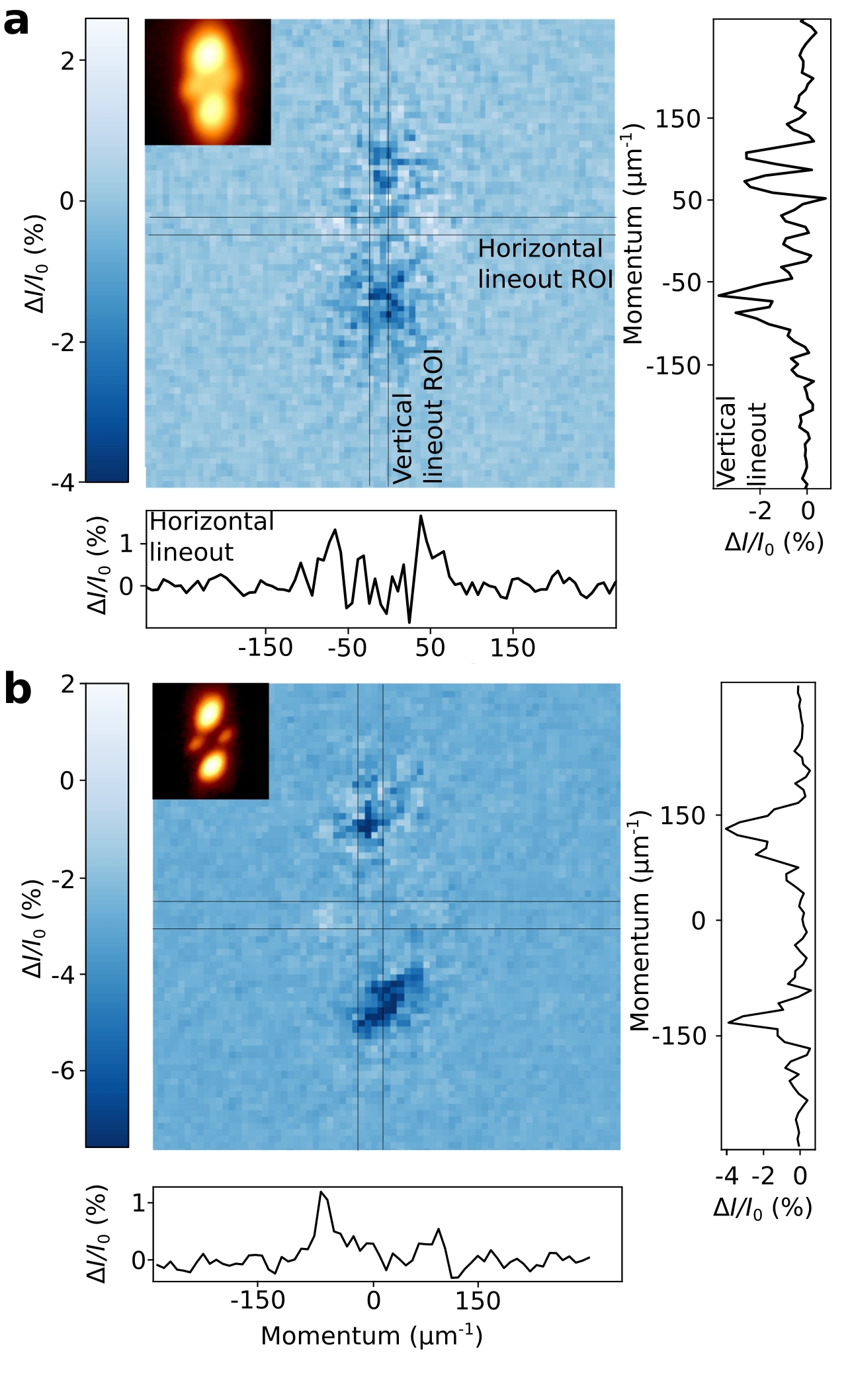}
\caption{Pump-probe diffraction snapshots of WSe$_2$/MoSe$_2$.
(a) 2$^\circ$   WSe$_2$/MoSe$_2$  at a 1 ps pump-probe delay, showing the change in diffraction intensity. Direct image is shown inset. (b) 57$^\circ$ WSe$_2$/MoSe$_2$ at a 1 ps pump-probe delay, showing the change in diffraction intensity. Direct image is shown inset.}
\end{figure}

\newpage

\begin{figure}[ht!]
\includegraphics[width=0.9\linewidth]{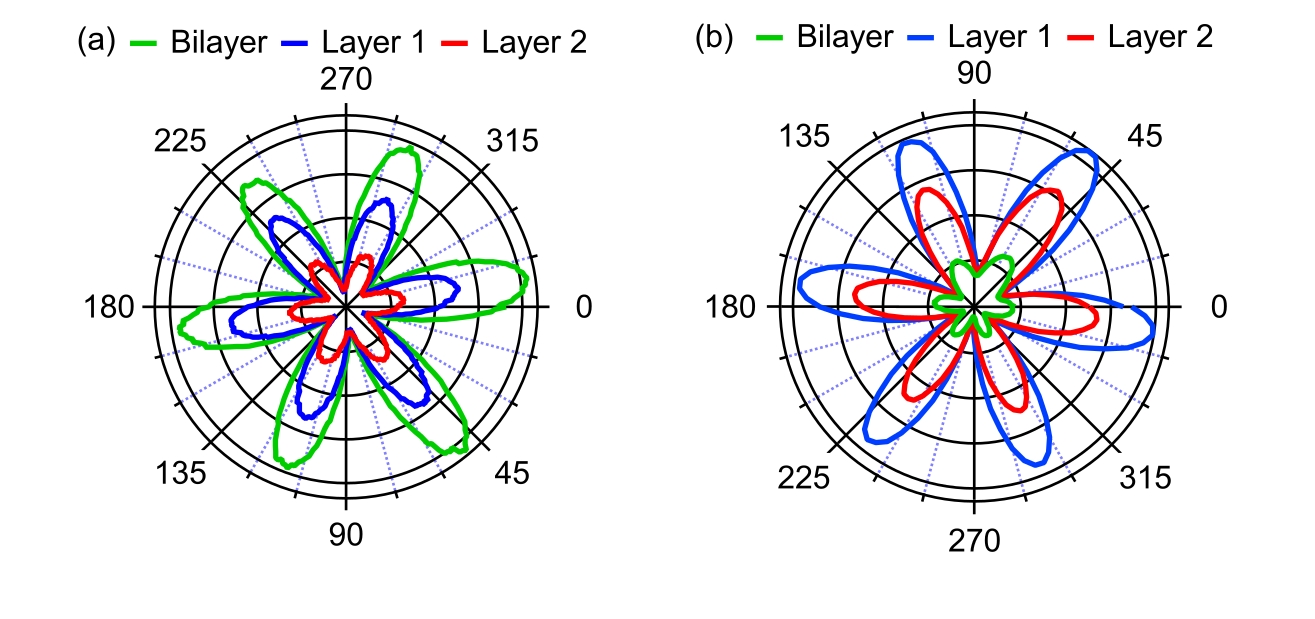}
\caption{ Second harmonic generation (SHG) polarization scan on the $2^\circ$ (a) and $57^\circ$ (b) WSe$_2$/MoSe$_2$ heterobilayers used in UED measurements. The alignment of crystal axes in the two layers are determined by constructive and destructive interference. The SHG signals are excited using a 1030 nm femtosecond laser with $<200$ fs pulse duration at room temperature, and captured by an EM CCD detector.}\end{figure} 

\newpage

\begin{figure}[ht!]
\includegraphics[width=0.8\linewidth]{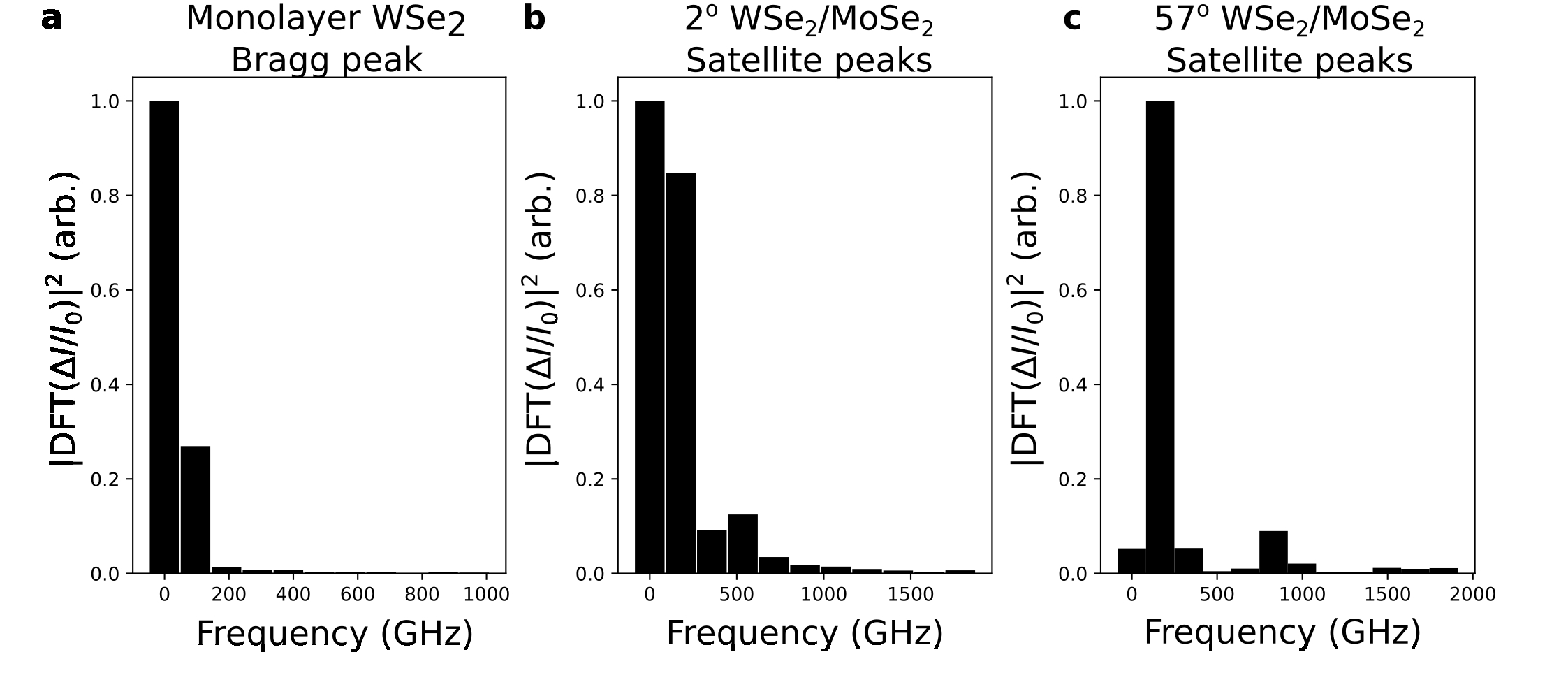}
\caption{ Power spectral density of UED time series.
Spectra are computed as the square modulus of the discrete Fourier transform (DFT) and normalized so that the maximum value is unity. No detrending is applied. (a) The DFT of the monolayer WSe$_2$ Bragg peak data (main-text Fig.~2(g)): the trend is monotonic decay with higher frequency, as predicted by a purely Debye-Waller (DW) thermal response. (b) The DFT of the $2^\circ$ WSe$_2$/MoSe$_2$ satellite peak data (main-text Fig.~2(a)): the trend contains two peaks, the DW contribution appearing at zero frequency and a second contribution that we attribute to the twisting-motion appearing in the bin $540 \pm 90$ GHz, overlapping with the central frequency of the best-fit dynamical model $510 \pm 90$ GHz. (c) The DFT of the $57^\circ$ WSe$_2$/MoSe$_2$ satellite peak data (main-text Fig.~2(d)): the DW contribution is peaked at  $160 \pm 80$ GHz, which is a windowing artifact indicating that the average change in diffraction intensity over all points in the time-scan is zero, i.e., the positive change mostly cancels the negative change. A second peak is present in the bin $830 \pm 80$ GHz, whereas the dynamical model of the experiment best fits $510 \pm 90$ GHz (the model spectrum is shown in main-text Fig.~2(f)): further discussion can be found in the Supplementary Information.}\end{figure}

\newpage

\begin{figure}[ht!]\centering
\includegraphics[width=0.8\linewidth]{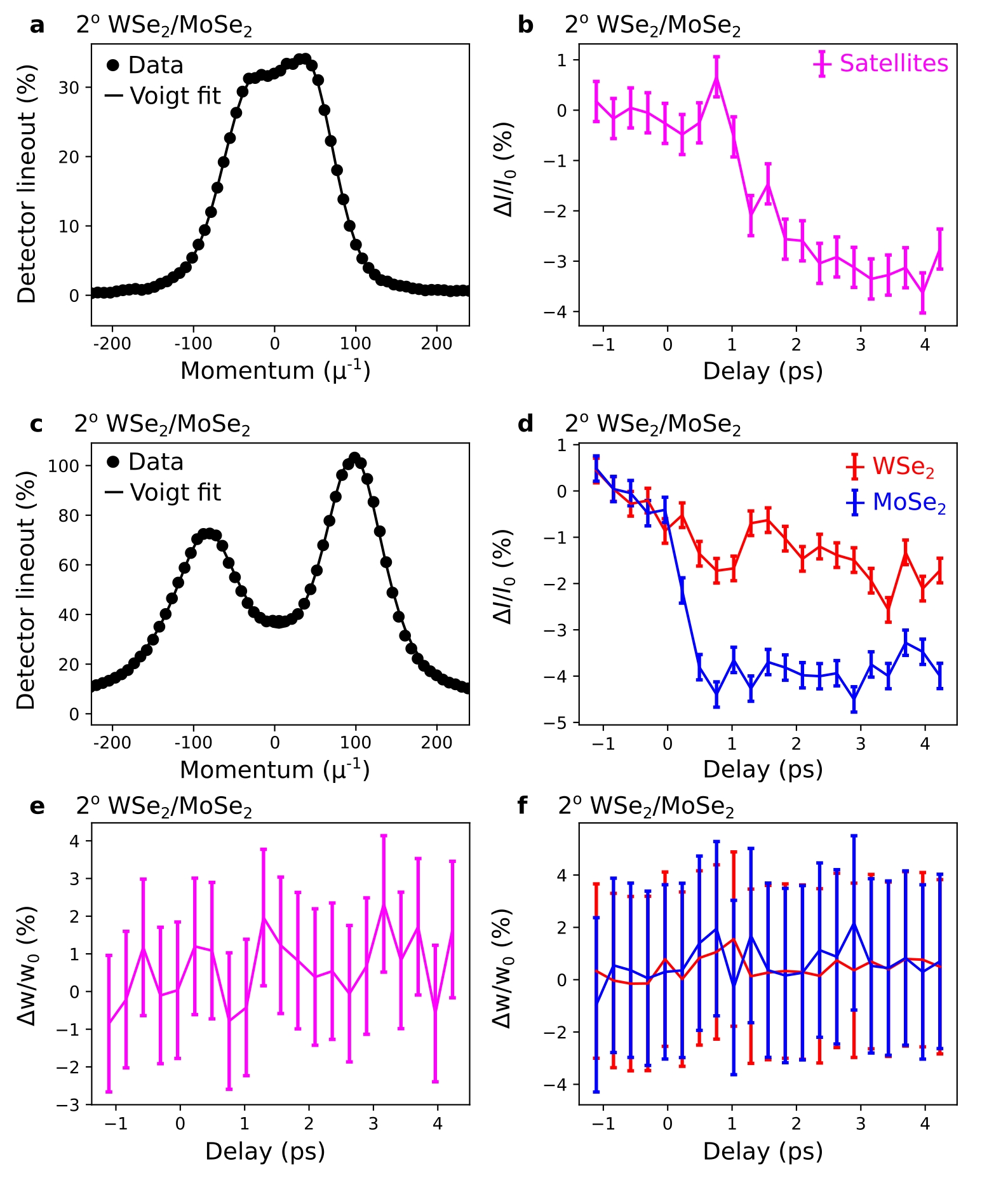}
\caption{ Voigt profile fits to $2{^\circ}$ WSe$_2$/ MoSe$_2$ UED data. 
Profiles are taken from lineouts marked in Extended Data~1(a), and indicate an alternative way to estimate the diffraction intensity compared to the method employed in the main text (summing counts within a region of interest). The choice of estimation technique does not change the main findings we report. (a) Example Voigt profile fit to satellite diffraction features. (b) Time series of Voigt-profile peaks fit to satellite diffraction features. Lines connecting data points are only a guide for the eye. (c) Example Voigt profile fit to atomic-scale Bragg diffraction features. (d) Time series of Voigt-profile peaks fit to atomic-scale Bragg diffraction features. Lines connecting data points are only a guide for the eye. (e)-(f) Time series of Voigt profile full-widths-at-half-maximum.}\end{figure}

\newpage

\begin{figure}[ht!] \centering
\includegraphics[width=0.8\linewidth]{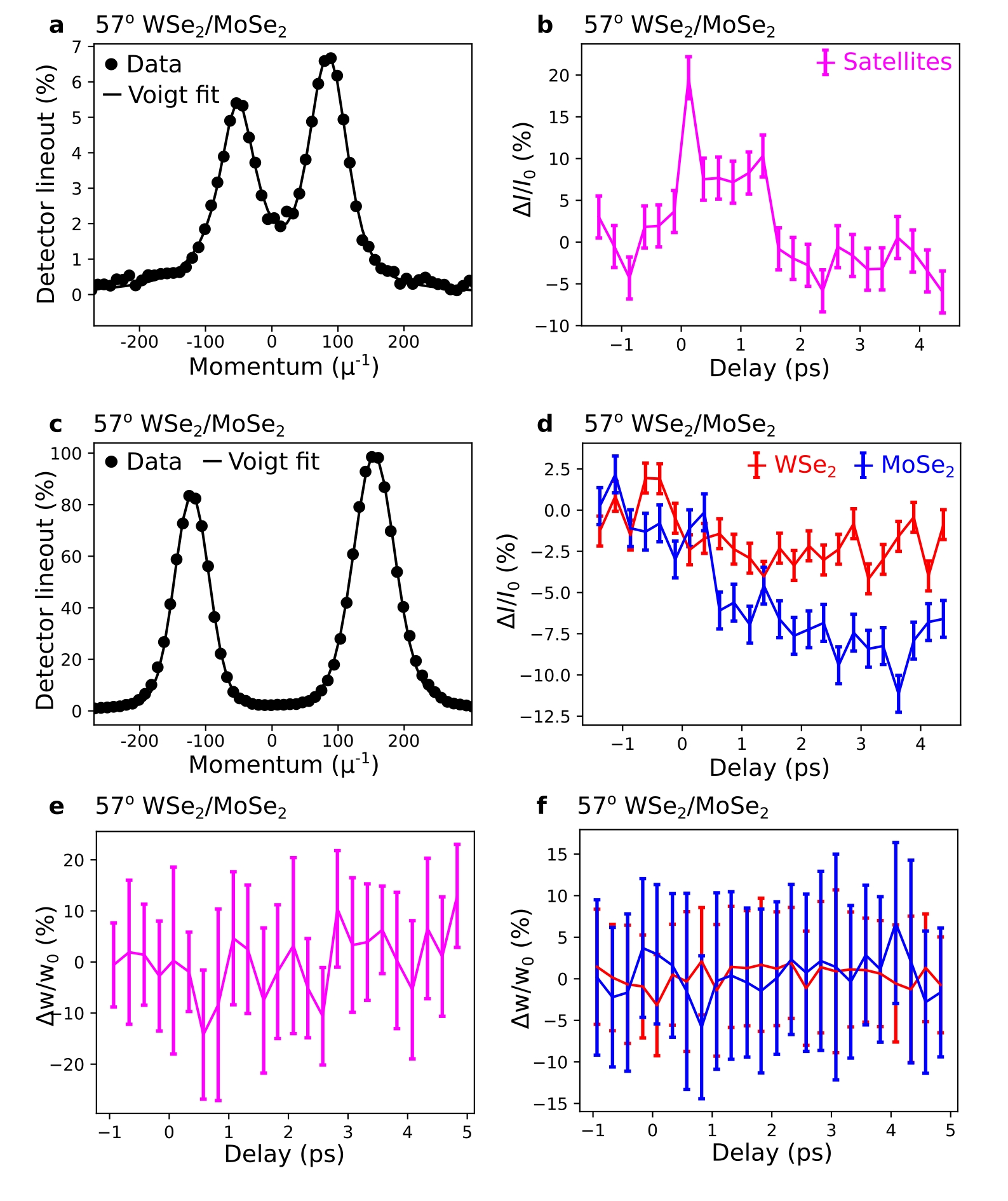}
\caption{
Voigt profile fits to $57{^\circ}$ WSe$_2$/ MoSe$_2$ UED data.
Profiles are taken from lineouts marked in Extended Data~1(b), and indicate an alternative way to estimate the diffraction intensity compared to the method employed in the main text (summing counts within a region of interest). The choice of estimation technique does not change the main findings we report.(a) Example Voigt profile fit to satellite diffraction features. (b) Time series of Voigt-profile peaks fit to satellite diffraction features. Lines connecting data points are only a guide for the eye. (c) Example Voigt profile fit to atomic-scale Bragg diffraction features. (d) Time series of Voigt-profile peaks fit to atomic-scale Bragg diffraction features. Lines connecting data points are only a guide for the eye. (e)-(f) Time series of Voigt profile full-widths-at-half-maximum.}\end{figure}

\newpage

\begin{figure}[ht!] \centering
\includegraphics[width=0.8\linewidth]{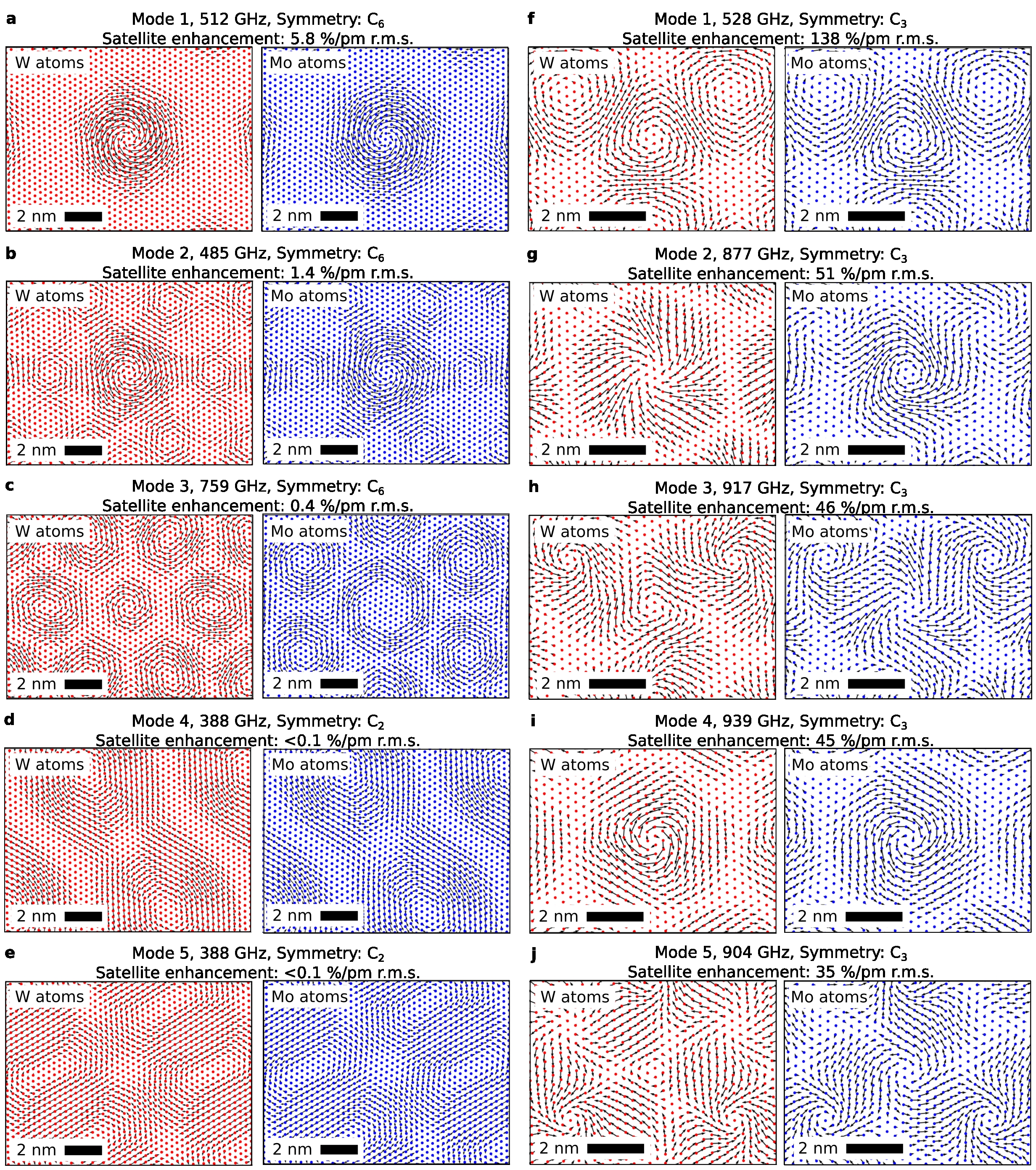}
\caption{ Diffraction sensitivity to lattice normal modes.
(a)--(e) Sensitivity of satellite peaks to lattice normal modes for 2$^\circ$ WSe$_2$ /MoSe$_2$. In this context, we define
the scattering enhancement $F$ to be the derivative of the peak intensity $I$ with respect to the r.m.s. atomic
displacement $x$ of the mode, normalized by the intensity $I_0$ at zero displacement: $F := (dI/dx)/I_0$ .
The top-five most efficient modes are ranked from most efficient (a) to least efficient (e). (f)--(j) Sensitivity of satellite peaks to lattice normal modes for 57$^\circ$~WSe$_2$/MoSe$_2$. The top-five most efficient modes are ranked from most efficient (f) to least efficient (j).}\end{figure} 

\newpage

\begin{figure}[ht!] \centering
\includegraphics[width=0.8\linewidth]{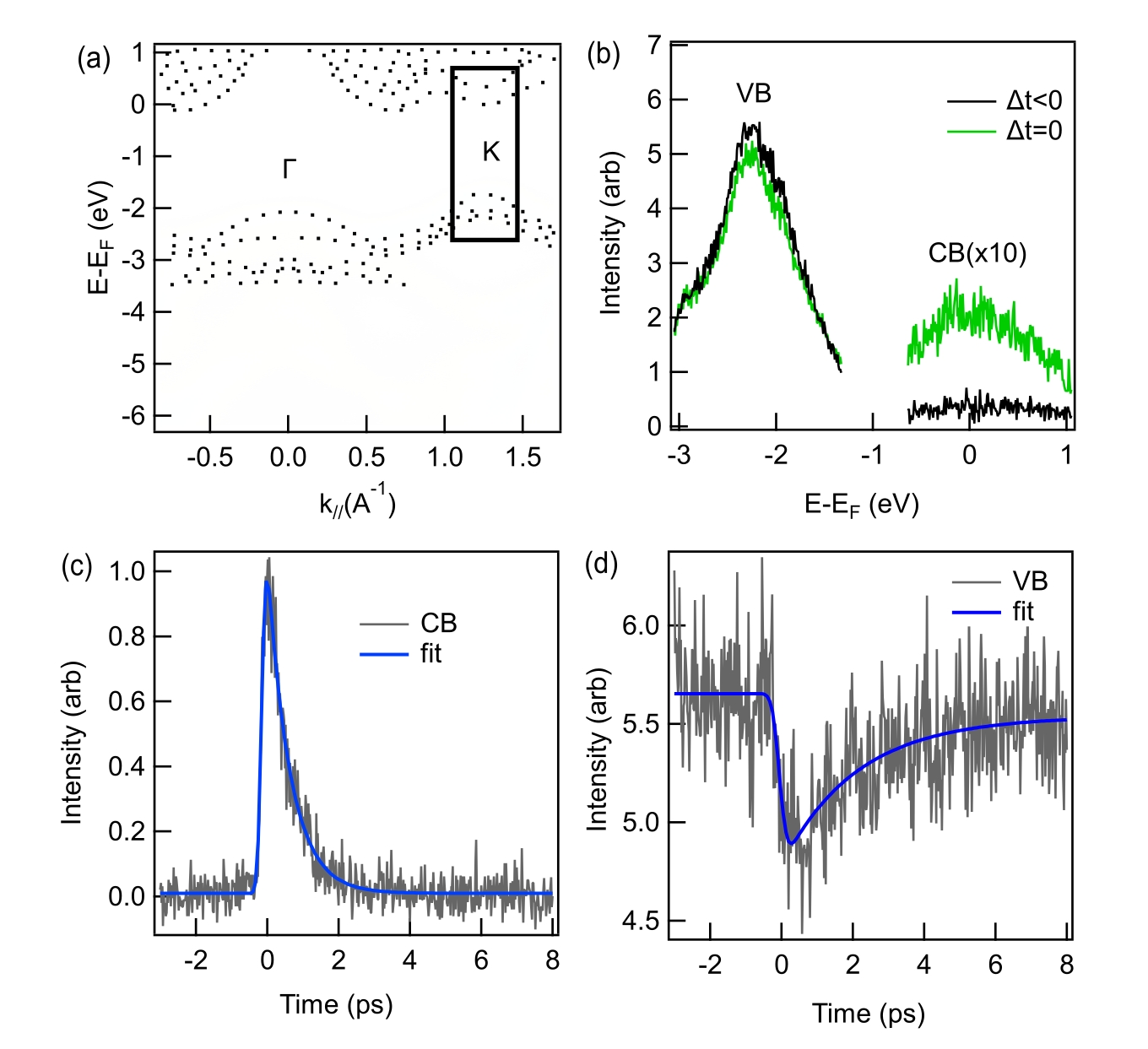}
\caption{
Time- and angle-resolved photoemission spectra 
(trARPES).
Measurements performed on a $3^\circ$ MoSe$_2/$WSe$_2$ 
heterostructure. (a) Static ARPES band structure. The 
dotted lines are theoretical prediction reproduced from 
\cite{GillenCalculation}. The box highlights the region 
around the K point at which the electrons and holes are 
monitored. (b) Electron energy distribution curve (EDC) 
near valence band (VB) and the conduction band (CB) 
edges at K point, before and after the pump excitation. 
(c)(d) Time-dependent CB and VB ARPES signals showing 
the evolution of electron and hole populations at the K 
point. Solid blue lines correspond to fit to exponential 
decay convoluted with the time resolution of $\sim$ 120 
fs. The CB electron dynamics is best fit to a single 
exponential decay with time constant of 630 fs. The 
population of the VB holes is best fit to a double 
exponential decay with time constants of 2 ps and 97 ps. 
The trARPES is performed with a 2.34 eV pump with a 
power density of $\sim$ 1.2 mJ/cm$^2$, and 21.7 eV EUV 
probe beam, under room temperature.}\end{figure}

\newpage

\begin{figure}[ht!] \centering
\includegraphics[width=0.65\linewidth]{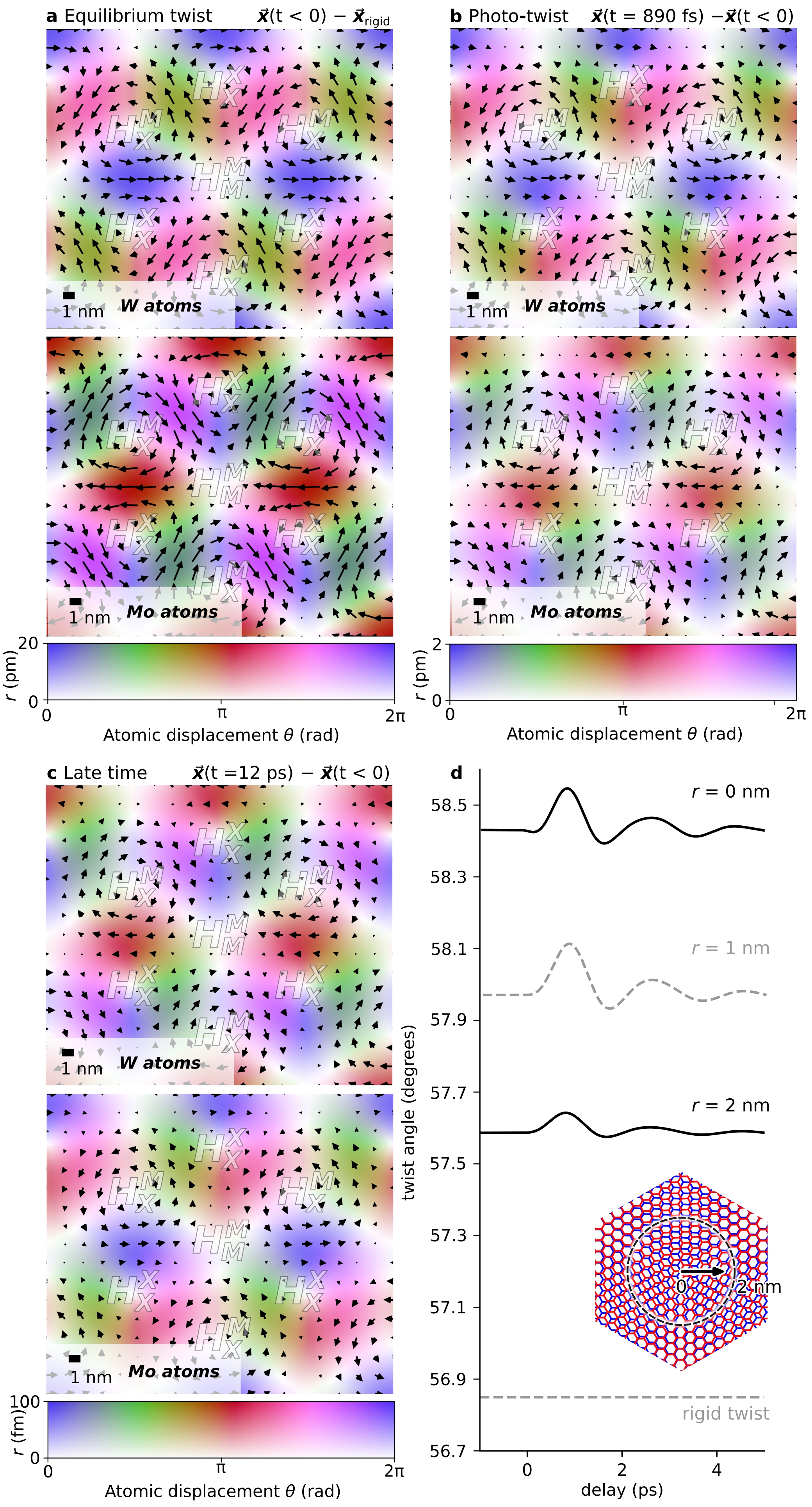}
\caption{
Twisting of the $57^\circ$ WSe$_2$/MoSe$_2$ lattice versus time.
Plots are extracted from our dynamical model, fit to  UED experimental data. (a) At equilibrium, atoms are displaced from from sites of the unphysical, rigidly rotated lattice by vdW forces. (b) Snapshot of the transverse displacement of atoms from equilibrium at 809 fs after photoexcitation. (c) Snapshot at 12 ps following photoexcitation, i.e., after the decay of the oscillatory transient. (d) Fitted atomic displacement as a function of radial distance from the vortex center, expressed as a twist angle.}\end{figure}

\newpage

\begin{figure}[ht!] \centering
\includegraphics[width=0.8\linewidth]{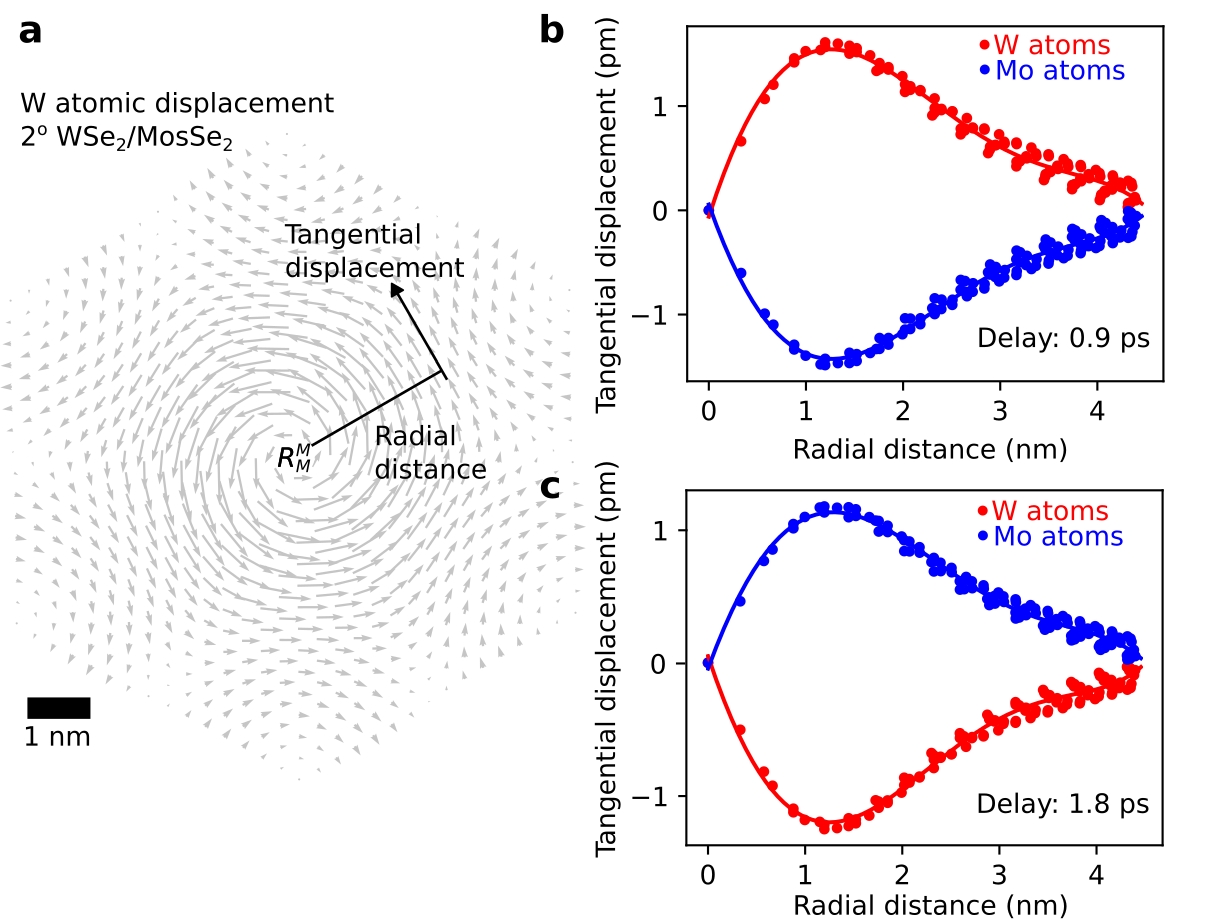}
\caption{
Procedure for extracting the local change in twist angle from dynamic simulations of atomic motion. (a) Definition of radial distance $r$ and tangential displacement $u$: local twist is defined to be $du/dr$ in the limit $r\rightarrow 0$. (b) Snapshot of the change in tangential displacement versus radial distance 0.9 ps after pumping the $2^\circ$ WSe$_2$/MoSe$_2$ sample, coinciding in time with the first peak of the oscillation. Dots show atomic positions, the solid line the best fifth order polynomial fit used to compute the derivate at $r=0$. Note that the tangential component of the displacement field is not single valued as a function of distance from the vortex center because the displacement field has only three-fold discrete rotational symmetry, rather than continuous rotational symmetry. The variance increases the greater the radial distance and there is consequently no improvement to the goodness of fit at polynomial orders higher than fifth. (c) A second snapshot 1.8 ps following pump arrival, coinciding with the first trough in the oscillation.}\end{figure}

\newpage

\begin{figure}[ht!] \centering
\includegraphics[width=0.8\linewidth]{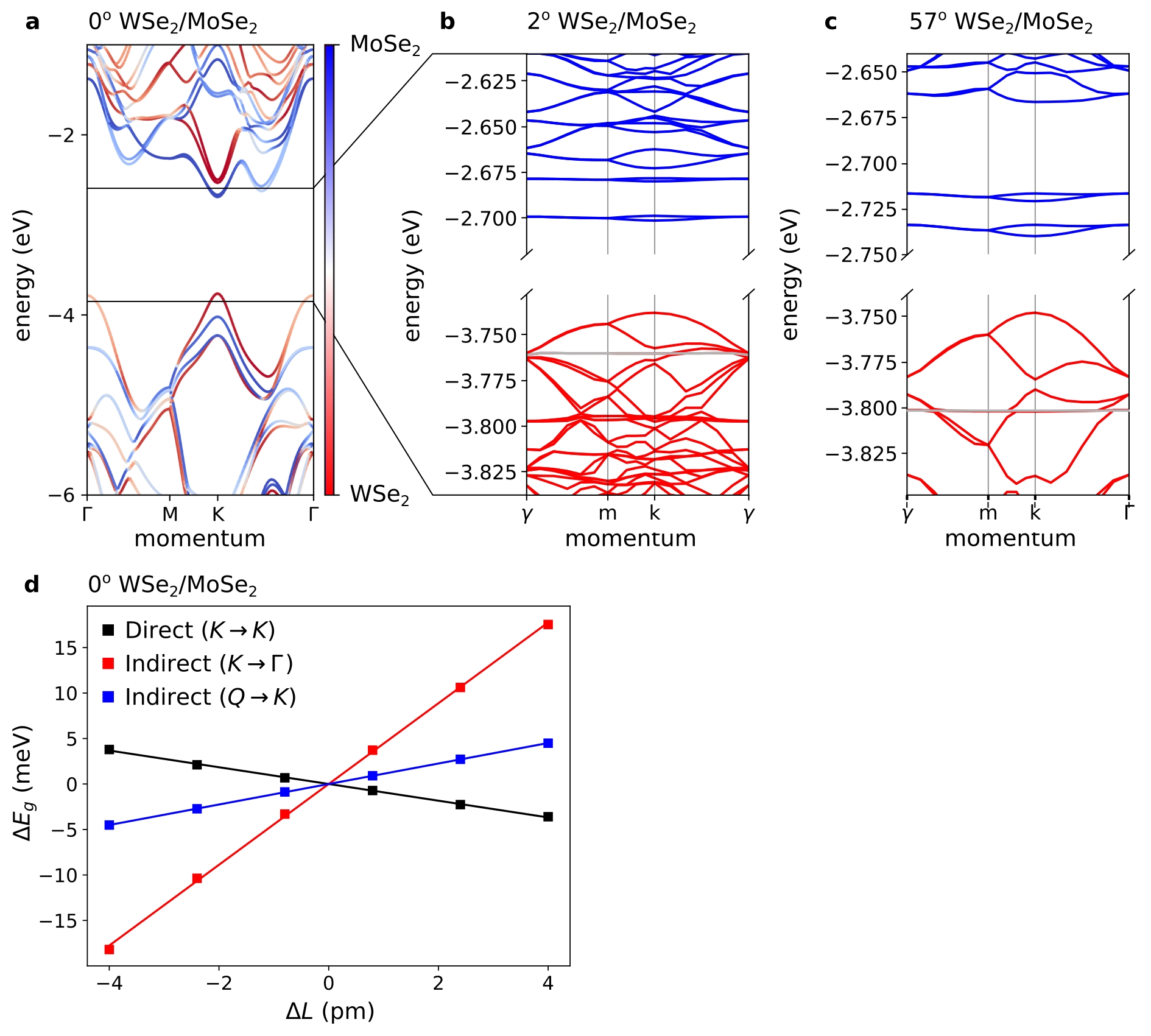}
\caption{
Simulated electron bands in WSe$_2$/MoSe$_2$.
Density-functional-theory calculations are performed according to the method summarized in the Supplemental Information. (a) $0^\circ$ twist angle, color map indicates whether the electron wavefunction is mostly localized to the WSe$_2$ layer (red) or MoSe$_2$ layer (blue). (b) $2.1^\circ$ twist angle showing electron bands over the moir\'{e} mini-Brillouin zone, blue bands are mostly localized to the MoSe$_2$ layer, red bands the WSe$_2$ layer. (c) $57^\circ$ twist angle, colors have the same meaning as in (b). (d) Change in bandgap $\Delta E_g$ as a function of change in interlayer spacing $\Delta L$ for momentum direct ($K\rightarrow K$) and momentum indirect ($K\rightarrow\Gamma, \ Q\rightarrow K$) transitions. The slope of the blue linear fit to $Q\rightarrow K$ is $1.1$ eV/nm, the value used to estimate the deformation-potential pressure in the main text.}\end{figure}

\end{document}